\documentclass[twocol]{ametsoc}

\journal{jcli}

\bibpunct{(}{)}{;}{a}{}{,}

\title{Climatology of severe local storm environments and synoptic-scale features over North America in ERA5 reanalysis and CAM6 simulation}

    \authors{Funing Li\correspondingauthor{Funing Li, 
    Purdue University, Department of Earth, Atmospheric, and Planetary Sciences, 550 Stadium Mall Drive, West Lafayette, IN 47907.}}
    \affiliation{Purdue University, Department of Earth, Atmospheric, and Planetary Sciences, West Lafayette, IN}
    \email{li3226@purdue.edu}
    
    \extraauthor{Daniel R. Chavas}
    \extraaffil{Purdue University, Department of Earth, Atmospheric, and Planetary Sciences, West Lafayette, IN}

    \extraauthor{Kevin A. Reed}
    \extraaffil{School of Marine and Atmospheric Sciences, Stony Brook University, Stony Brook, NY}
    
    \extraauthor{ Daniel T. Dawson II}
    \extraaffil{Purdue University, Department of Earth, Atmospheric, and Planetary Sciences, West Lafayette, IN}
	
\abstract{Severe local storm (SLS) activity is known to occur within specific thermodynamic and kinematic environments. These environments are commonly associated with key synoptic-scale features--including southerly Great Plains low-level jets, drylines, elevated mixed layers, and extratropical cyclones--that link the large-scale climate to SLS environments. This work analyzes spatiotemporal distributions of both the environmental parameters and synoptic-scale features in ERA5 reanalysis and in Community Atmosphere Model version 6 (CAM6) during 1980--2014 over North America. Compared to radiosondes, ERA5 successfully reproduces SLS environments, with strong spatiotemporal correlations and low biases, especially over the Great Plains. Both ERA5 and CAM6 reproduce the climatology of SLS environments over the central United States as well as its strong seasonal and diurnal cycles. ERA5 and CAM6 also reproduce the climatological occurrence of the synoptic-scale features, with the distribution pattern similar to that of SLS environments. Compared to ERA5, CAM6 exhibits a high bias in Convective Available Potential Energy over the eastern United States primarily due to a high bias in surface moisture, and to a lesser extent, storm-relative helicity due to enhanced low-level winds. Composite analysis indicates consistent synoptic anomaly patterns favorable for significant SLS environments over much of the eastern half of the United States in both ERA5 and CAM6, though the pattern differs for the southeastern United States. Overall, results indicate that both ERA5 and CAM6 are capable of reproducing SLS environments as well as the synoptic-scale features and transient events that generate them. }

\begin{document}

\maketitle


\section{Introduction}\label{sec:introduction}

Severe local storm (SLS) environments are favorable atmospheric conditions for the development of SLS events, including severe thunderstorms accompanied by damaging winds, large hailstones, and/or tornadoes \citep{ludlam1963, johns1992}. Such environments are commonly defined by high values of a small number of key thermodynamic and kinematic parameters: convective available potential energy (CAPE), lower-tropospheric (0--6-km) bulk vertical wind shear (S06), and 0--3-km storm-relative helicity (SRH03) \citep{Rasmussen_Blanchard_1998, Rasmussen_2003, Brooks_etal_2003, Doswell_Schultz_2006, Grams_etal_2012}. CAPE is the vertical integral of buoyancy from the level of free convection to the equilibrium level and thus provides a measure of conditional instability for a potential storm \citep{Doswell_Rasmussen_1994}. Given the standard assumptions of parcel theory, CAPE is proportional to the theoretical maximum updraft wind speed \citep{holton_1973}. S06 and SRH03 are proxies of environmental crosswise and streamwise vorticity available to generate updraft rotation \citep{rotunno_1982, rotunno_1985, Davies_1990, Davies_1993, Weisman_Rotunno_2000, Davies-Jones2002, Rotunno2003, Davies-Jones2003}. These conditions work in combination to permit the generation of persistent rotating updrafts that are the defining characteristic of supercell storms \citep{Doswell_Burgess_1993}. Given that both thermodynamic and kinematic ``ingredients'' are necessary, composite proxies, such as the product of CAPE and S06 (CAPES06) and the energy helicity index (EHI03; proportional to the product of CAPE and SRH03), are commonly employed as representative measures of the SLS potential of a given environment \citep{Davies_1993,Brooks_etal_2003, thompson_2003}. Composite proxies are in general preferable to the constituent parameters alone in discriminating conducive environments for SLS events \citep{Rasmussen_Blanchard_1998, Brooks_etal_2003}, while the individual constituent parameters are indicative of the key underlying physical processes \citep{Doswell_Schultz_2006}. 

The generation of SLS environments depends on synoptic-scale features that serve as an intermediate-scale bridge between weather and climate scales. Over the Great Plains of North America, the generation of SLS environments is intimately associated with the southerly Great Plains low-level jets, drylines, elevated mixed layers, and extratropical cyclones. The southerly Great Plains low-level jets (GPLLJs), defined by the low-level maximum winds, commonly form during the nighttime \citep{Bonner_1968, Whiteman_etal_1997}. GPLLJs modulate Great Plains precipitation \citep{weaver2008} and its moisture budget by transporting almost one-third of the moisture that enters the contiguous United States \citep{Helfand_Schubert_1995}, which is an important contributor to the formation of high-CAPE environments \citep{Helfand_Schubert_1995, Higgins_etal_1997, weaver2012}. Drylines and elevated mixed layers (EMLs) are associated with the eastward advection of well-mixed air with relatively high potential temperature generated by strong surface heating over the elevated dry deserts of the Mexican Plateau. Meridionally oriented drylines are common across the central and southern Great Plains when this dry airmass from the west encounters the moist near-surface airmass advected northward from the Gulf of Mexico \citep{Fujita_1958, Schaefer_1974, Ziegler_Hane_1993, Hoch_Markowski_2005}, producing a focused linear band of convergence and moisture gradients \citep{ Rhea_1966, Ziegler_Rasmussen_1998, Xue_Martin_2006, Schultz_etal_2007}. The advection of the well-mixed layer atop the low-level moist air generates the EML, often creating a strong capping inversion at the base of the EML \citep{Carlson_etal_1983,Lanicci_Warner_1991a,Banacos_Ekster_2010}. Subsequent daytime heating of the moist boundary layer allows for the removal of convective inhibition and a strong buildup of CAPE \citep{Carlson_etal_1983,Farrell_Carlson_1989, Lanicci_Warner_1991b,Lanicci_Warner_1991c, Cordeira_etal_2017}. Extratropical cyclones strongly enhance low-level moisture and heat convergence within their warm sectors, thereby enhancing CAPE \citep{Hamill_etal_2005, Tochimoto_etal_2015}. Moreover, the cyclonic circulation itself and the baroclinic instability linked to the evolution of extratropical cyclones via thermal wind balance also act to enhance the vertical wind shear \citep{Doswell_Bosart_2001}. 

Previous studies have documented the climatological variability, including the amplitude and spatial pattern, of the SLS environmental proxies and parameters \citep{Brooks_etal_2003, Wagner_etal_2008, gensini_ashley_2011, Diffenbaugh_etal_2013, tippett_2015, Gensini_Brooks_2018, gensini_lelys_2019, tang2019}, as well as the key synoptic-scale features that help generate these environments \citep{Bonner_1968, reitan_1974, zishka_1980, Lanicci_Warner_1991a, Whiteman_etal_1997, Hoch_Markowski_2005, Duell_Broeke_2016, Ribeiro_Bosart_2018}. It is known from these studies that the SLS environments and the associated synoptic-scale features have strong seasonal and diurnal cycles, which emphasizes the influence of variations in the fundamental types of external climate forcing (earth orbit and solar insolation). However, these studies have analyzed different geographic domains and time-periods using different datasets, which makes holistic analysis and inter-comparison difficult.

Reanalysis datasets, combining multi-source observations and model-based forecasts via advanced data assimilation methods, provide a synthesized representation of the atmosphere over long historical time periods and thus have become an indispensable data source for the study of weather and climate from synoptic to planetary scales. Several reanalysis datasets have been used to investigate the climatological distribution or variation of SLS environments, including NCEP/NCAR \citep{Brooks_etal_2003, Diffenbaugh_etal_2013}, North American Regional Reanalysis (NARR) \citep{gensini_ashley_2011, Gensini_etal_2014, Tippett_etal_2016, Gensini_Brooks_2018,  tang2019}, and ERA-Interim \citep{allen2014,Taszarek_etal_2018}. These reanalyses in general produce similar spatial patterns but slightly different magnitudes for the climatology of SLS environments. Recent work evaluated the performance of ERA-Interim and NARR in estimating the environmental proxies and parameters by comparing them with radiosonde measurements \citep{Gensini_etal_2014, Taszarek_etal_2018}, indicating that these reanalyses reproduce the observed spatial and temporal trends of SLS environments, though they exhibit larger regional biases in thermodynamic parameters (e.g., CAPE) than in kinematic parameters (e.g., S06). In addition, these reanalyses in general produce lower biases over flat terrain than in coastal areas and mountains owing to the limited horizontal resolution and sharp variations of atmospheric variables across regions with complex topography \citep{Taszarek_etal_2018}. Thus, caution is needed when interpreting results from reanalysis datasets considering the various uncertainties associated with the measurements and forecasts incorporated, as well as the complex inferential process that involves data assimilation for generating reanalysis datasets \citep{parker_2016}. Such evaluation is necessary for identifying potential strengths and limitations of a reanalysis dataset for studying SLS environments \citep{Gensini_etal_2014} and a higher-resolution reanalysis dataset is expected to better represent these atmospheric environments. 

Additionally, global climate models have been a useful tool to study SLS environments \citep{tippett_2015}. As with reanalyses, such models are too low-resolution to resolve actual SLS events, but they are capable of resolving larger-scale SLS environments. Though climate models do not reproduce the day-to-day weather, they are able to capture the statistical behavior of the present-day climate. Thereby, climate models have been widely used to analyze SLS environments in current or future climate simulations to assess the impacts of climate changes on SLS activity, assuming that changes in SLS activity will follow changes in the statistics of SLS environments \citep{Trapp_etal_2007, Diffenbaugh_etal_2013, Romps_etal_2014, Seeley_Romps_2015,Tippett_etal_2016, Gensini_Brooks_2018}. These models in general reproduce reasonable historical climatologies of SLS environments, though biases vary across models when compared to the reanalysis- or radiosonde-based climatology \citep{Diffenbaugh_etal_2013, Seeley_Romps_2015}. Forced with coarse-resolution output from global climate models, dynamical downscaling through high-resolution regional climate models has shown substantial potential for assessment of both SLS events and environments \citep{Trapp_etal_2011, Robinson_etal_2013, Gensini_Mote_2014, Gensini_Mote_2015, Hoogewind_etal_2017}.

The purpose of this study is therefore to evaluate the representations of both SLS environments and synoptic-scale features commonly associated with the generation of these environments over North America in a new high-resolution global reanalysis dataset (ERA5) and a climate model (the Community Atmosphere Model version 6, CAM6). A comprehensive analysis and comparison of the climatologies of these environments and the synoptic-scale features provides an important reference for using reanalyses and climate models to better understand climate controls on SLS activities in any climate state. 

This work addresses the following questions:

\begin{enumerate}
\item How well does the ERA5 reanalysis represent the observed climatology of SLS environments over the contiguous United States?
\item How does this climatology, including seasonal and diurnal cycles, compare between the ERA5 reanalysis and CAM6 simulation?
\item What are the climatological distributions of the key synoptic-scale features commonly associated with the generation of SLS environments over North America in the ERA5 reanalysis and CAM6 simulation?
\item What are the characteristic synoptic patterns associated with SLS environments in the ERA5 reanalysis and CAM6 simulation? Do they vary from region to region?
\end{enumerate}

To answer these questions, we first compare SLS environments between the ERA5 reanalysis and radiosonde observations. This examines the ability of the ERA5 reanalysis in reproducing both statistical property of the present-day climate and the observed day-to-day weather in terms of SLS environments. As climate models simulate climate statistics, we then compare the climatology, including seasonal and diurnal cycles, of extreme SLS environments and the associated synoptic-scale features between the ERA5 reanalysis and CAM6 simulation, as well as analyze biases in the CAM6 simulation. Finally, we create and compare synoptic composites associated with extreme SLS environments within a set of predefined regions across the eastern half of the United States to analyze the extent to which CAM6 reproduces the characteristic synoptic-scale flow patterns that generate these events across regions.

Section \ref{sec:methdology} introduced the data, experimental design, and our analysis methodology. We evaluate the ERA5 reanalysis against radiosonde observations in Section \ref{sec:results_era5}. Section \ref{sec:results_ctrl} presents climatologies of SLS environments and synoptic-scale features, and synoptic composites. Finally, we provide conclusions and discussion of results and future work in Section \ref{sec:conclusions}. 

\section{Methodology}\label{sec:methdology}

\subsection{Datasets}

We use radiosonde observations for the period 1998--2014, and the ERA5 reanalysis and CAM6 historical simulation data each for the period 1980--2014. The radiosonde observations are first used to evaluate the representation of SLS environments in ERA5 over the contiguous United States. Then, the CAM6 simulation is compared in-depth to ERA5 for North America. Each data source is described in detail below.

\subsubsection{Observation: Radiosonde}

Radiosonde observations are obtained from the sounding database of the University of Wyoming
(\url{http://weather.uwyo.edu/upperair/sounding.html}). This dataset includes 69 radiosonde stations over the contiguous United States with twice-daily raw soundings at 0000 and 1200 UTC. Three stations (the KUNR station over western South Dakota, the KVEF station over southern Nevada, and the KEYW station on the island of Key West) are excluded in this study because of a lack of multi-year records, resulting in 66 stations (Figure \ref{fig_map}a). Roughly half of the radiosonde stations from the database do not have records before around 1994 and most stations were moved 0.01--0.03 degree along the longitude or latitude in 1990's (in or before 1997) due to the National Weather Service modernization. Thus, we only use radiosonde observations for the period 1998--2014 in this work, similar to \citet{Gensini_etal_2014}. Furthermore, we apply the following quality-control checks to each sounding before use: (1) height and pressure arrays are in correct order: height increases and pressure decreases with time; (2) wind speed $\geq$ 0 kts, 0$^\circ$ $\leq$ wind direction $\leq$ 360$^\circ$, and temperature and dewpoint temperature $\geq$ 0 K; (3) height of the first record equals the local elevation; (4) top height $>$ 6 km and top pressure $<$100 hPa; and (5) the maximum pressure decrease between consecutive records $\leq$ 50 hPa. The first two checks follow the quality control done in SHARPpy \citep{Blumberg_etal_2017}; the third check ensures that the ground surface observation is available for calculating the surface-based CAPE; the fourth and fifth checks ensure that the vertical resolution of the sounding is identical with or higher than the ERA5 reanalysis, as the purpose of using radiosondes is to evaluate the ERA5 representations. The number percentage of qualified records for each radiosonde station is over 60\% (Figure S1a).

\subsubsection{Reanalysis: ERA5}

The fifth generation of the European Centre for Medium-Range Weather Forecasts (ECMWF) global climate reanalysis, ERA5, spans the period 1979--present \citep{hersbach_2016}. Here we use years 1980--2014, downloaded from NCAR's Research Data Archive \cite[date accessed: 09-19-2019;][]{rda_ds633.0}, for direct comparison with radiosondes and climate model simulation (described below). The dataset provides hourly variables at or near the surface and 37 constant pressure levels from 1000--1 hPa. These pressure-level data are produced by interpolating from the ECMWF's Integrated Forecast System with 137 hybrid sigma-pressure model levels in the vertical, up to a top level of 0.01 hPa \citep{hersbach_2016}; this reduction of vertical resolution may induce errors to calculations of vertically integrated parameters, such as CAPE and SRH03 (defined below). The horizontal grid spacing of ERA5 is 0.25 degree (roughly 31 km), which is higher than its predecessor ERA-Interim \cite[79 km;][]{dee_2011}. Other improvements in ERA5 include using a revised data assimilation system and improved core dynamics and model physics \citep{hersbach_2016}. 

\subsubsection{Modeling: CAM6}

The Community Atmosphere Model version 6 (CAM6) is used for the simulation portion of this work. CAM6 is the atmospheric component of the Community Earth System Model version 2.1 ( available at \url{ http://www.cesm.ucar.edu/models/cesm2/}) developed in part for participation in the Coupled Model Intercomparison Project 6 \cite[CMIP6;][]{eyring2016}. CAM6 builds off its predecessor CAM5 \cite[documented in detail in][]{Neale_etal_2012} with significant modifications to the physical parameterization suite. In particular, CAM5 schemes for cloud macrophysics, boundary layer turbulence and shallow convection have been replaced by the Cloud Layers Unified by Binormals \cite[CLUBB;][]{golaz2002,bogenschutz2013} scheme. In addition, CAM6 now implements the two-moment prognostic cloud microphysics from \citet{gettelman2015}, as well as additional updates to the \citet{zhang1995} deep convection and orographic drag parameterizations. CAM6 is configured with the default finite volume dynamical core on a 0.9$^\circ\times$1.25$^\circ$ latitude-longitude grid mesh with 32 hybrid sigma-pressure levels. Our CAM6 simulation is configured as a historical simulation following Atmospheric Model Intercomparison protocols \citep{Gates_etal_1999} over the period 1979--2014. We discard the first year for spinup and analyze the 3-hourly output from 1980--2014 for direct comparison with ERA5 reanalysis. 

\begin{figure}[t]
\centerline{\includegraphics[width=19pc]{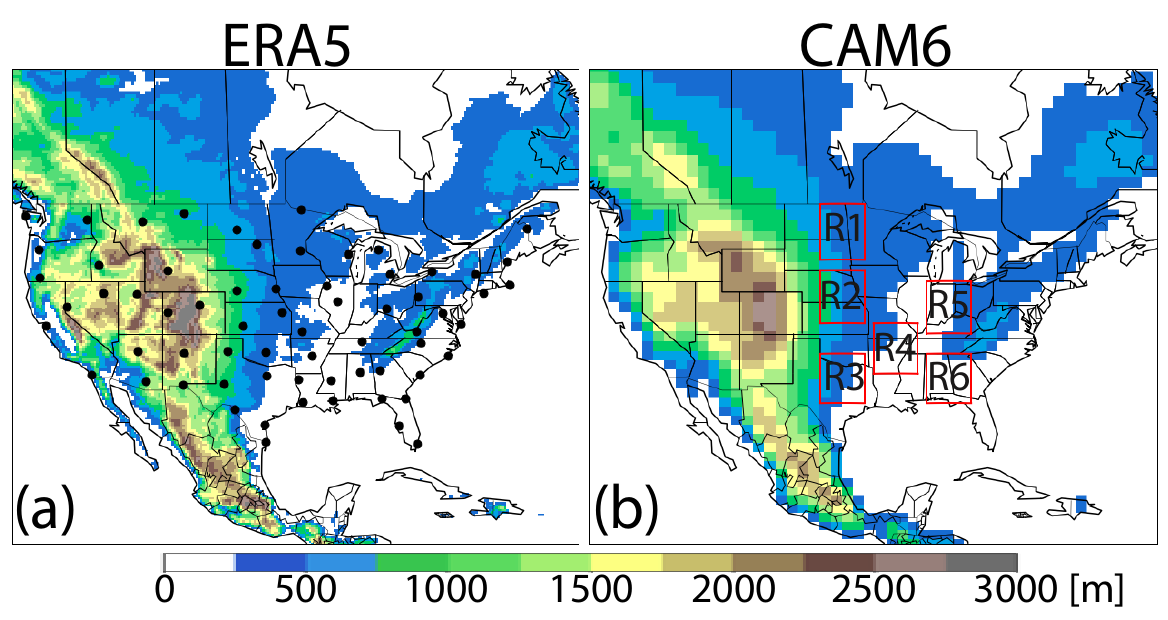}}
\caption{Elevation map for (a) ERA5 reanalysis and (b) CAM6 simulation over North America. Black dots indicate locations of the 66 radiosonde stations over the contiguous United States. R1--R6 denote the six 5$^\circ$$\times$$5^\circ$ sub-region boxes selected for regional analysis.}
\label{fig_map}
\end{figure}

\subsection{Analysis}

We perform a climatological analysis of: (1) the annual, seasonal, and diurnal distributions of significant SLS environmental proxies and their constituent parameters (defined below); (2) the occurrence frequency distributions of key synoptic-scale features: southerly Great Plains low-level jets, drylines, elevated mixed layers, and extratropical cyclone activity over North America; and (3) characteristic synoptic composites associated with extreme SLS environments in different geographic regions over the eastern half of the United States. To evaluate ERA5 performance using radiosondes, we extract ERA5 at 0000 and 1200 UTC for 1998--2014, consistent with the radiosonde temporal resolution and coverage, and then linearly interpolate ERA5 results onto the radiosonde sites. To compare simulation with reanalysis, calculations for ERA5 and CAM6 are all done for 3-hourly outputs from 1980--2014 and then linearly interpolated onto 1$^\circ$$\times$1$^\circ$ grids. These horizontally linear interpolations are performed via the function of $matplotlib.mlab.griddata$ in the open-source \textit{matplotlib} Python package \citep{Hunter2007}.

\subsubsection{SLS Environments}

\textbf{SLS Environmental Proxies and Parameters:} We calculate two combined proxies, CAPES06 and EHI03, to represent SLS environments. CAPES06 \citep{Brooks_etal_2003} and EHI03 \citep{Hart_Korotky_1991, Davies_1993} are calculated by 
\begin{equation} \label{eq_capes06} {CAPES06 = CAPE\times S06}, \end{equation} and 
\begin{equation} \label{eq_ehi} {EHI03 = \frac{CAPE\times SRH03}{160,000\ m^{4}\ s^{-4}}}, \end{equation} respectively.
As for each constituent parameter, CAPE is defined as \citep{Doswell_Rasmussen_1994}:
\begin{equation} \label{eq_cape} CAPE={\int_{z_{LFC}}^{z_{EL}} g\frac{T_{vp}-T_{ve}}{T_{ve}} dz}, \end{equation}
where $g=$ 9.81 m s$^{-2}$ is the acceleration due to gravity, $z_{LFC}$ denotes the level of free convection, $z_{EL}$ denotes the equilibrium level, and $T_{vp}$ and $T_{ve}$ is the virtual temperature of the 2-m parcel and the environment. Here we select the 2-m parcel for simplicity its consistent availability across all our datasets; it also avoids biases in defining other types of parcels (e.g., most-unstable or mixed-layer parcel) associated with differences in the vertical resolutions of the datasets. S06 is defined as the magnitude difference of wind vectors at 6 km and 10 m above the surface \citep{Rasmussen_Blanchard_1998, Weisman_Rotunno_2000}. SRH03 is defined as \citep{Davies_1990} \begin{equation} \label{eq_srh} SRH03={-\int_{z_b}^{z_t} \hat{\textbf{k}}\cdot(\textbf{V}-\textbf{C})\times\frac{\partial\textbf{V}}{\partial{z}} dz}, \end{equation}
where \textbf{V} is horizontal wind vector, \textbf{C} is the storm motion vector following the definition and calculation from \citet{bunkers_2000}, $z_b=10$ m is the altitude of the layer bottom, $z_t=3$ km is the altitude of the layer top, and $\hat{\textbf{k}}$ is the vertical unit vector. We create the climatologies of SLS environments using CAPES06 and EHI03, respectively, as well as their constituent parameters (CAPE, S06, and SRH03). 

\textbf{Analysis of Extremes:} We define ``significant'' (or ``extreme'') SLS environments using the 99th percentile of the proxies and parameters, similar to past work \citep{Tippett_etal_2016, Singh_etal_2017}. The 99th percentile is calculated for each station site (for radiosondes) or each grid point (for ERA5 reanalysis or CAM6 simulation) based on the full-period (1998--2014 or 1980--2014) time series of each variable at the location. We also analyzed the 95th, 90th, and 75th percentiles, and found qualitatively similar climatological patterns across these high percentiles (analyzed below in Figure S4). 

\subsubsection{SLS-relevant synoptic-scale features}

\textbf{Southerly Great Plains Low-Level Jet (GPLLJ):} We follow \citet{Bonner_1968} and \citet{Whiteman_etal_1997} to identify low level jets (LLJs). The method detects LLJs and defines an intensity category 0--3 based on two criteria: (1) the maximum wind speed below 3000 m: $V_{max}\geq$ 10, 12, 16, or 20 m s$^{-1}$ and (2) the largest decrease from $V_{max}$ in the layer from the height of $V_{max}$ to 3000 m: $\Delta V\geq$ 5, 6, 8, or 10 m s$^{-1}$. To identify specifically southerly LLJs, each detected LLJ is further classified as southerly if the direction of $V_{max}$ falls between 113$^\circ$ and 247$^\circ$, and as northerly if between 293$^\circ$ and 67$^\circ$ following \citet{Walters_etal_2008} and \citet{Doubler_etal_2015}. Using these category- and direction-based criteria, we successfully detect various LLJs over North America, including the southerly and northerly GPLLJ, the northerly Pacific coast LLJ, the northerly Tehuantepec LLJ, and the easterly Caribbean LLJ, as documented in \citet{Doubler_etal_2015}. The climatology of southerly GPLLJ in category 0 ($V_{max}\geq$ 10 m s$^{-1}$ and $\Delta V\geq$ 5 m s$^{-1}$) is presented in this work.   

\textbf{Dryline:} Drylines are identified at each grid point following the criteria in \citet{Duell_Broeke_2016}: (1) the horizontal gradient of the surface specific humidity is at least 0.03 g kg$^{-1}$ km$^{-1}$ and the specific humidity gradient from west to east must be positive, (2) the surface temperature gradient from west to east is less than 0.02 K km$^{-1}$, and (3) a surface wind shift exists with wind direction on the west side being between 170$^\circ$ and 280$^\circ$, and on the east side being between 80$^\circ$ and 190$^\circ$. The first criterion is consistent with the approach of \citet{Hoch_Markowski_2005}, in which the specific humidity gradient is recommended instead of the dewpoint temperature gradient as the specific humidity is less sensitive to the varying elevation. The other two criteria are used in an effort to differentiate drylines from cold fronts. The limitations of the algorithm are further discussed in \citet{Duell_Broeke_2016}.

\textbf{Elevated Mixed Layer (EML):} An EML is identified for a sounding that satisfies the following criteria, based on \citet{Banacos_Ekster_2010} and \citet{Ribeiro_Bosart_2018}: (1) a candidate EML is identified as a layer with lapse rate equal to or greater than 8.0 K km$^{-1}$ through a depth of at least 200 hPa, (2) the environmental relative humidity increases from the base to the top of the candidate EML, (3) the base of the candidate EML is at least 1000 m above the surface but below the 500-hPa level, and (4) the average lapse rate between the base and the surface is less than 8.0 K km$^{-1}$. Here, the EML base is defined as the first model level from the bottom with lapse rate equal to or greater than 8.0 K km$^{-1}$. The third and fourth criteria ensure the exclusion of surface-based or upper-tropospheric mixed layers.
 
\textbf{Extratropical Cyclone Activity:} Extratropical cyclone activity is defined using two methods: (1) cyclone track frequency calculated from explicit tracking of cyclone centers, and (2) eddy kinetic energy (EKE), which captures the spatial distribution of eddy activity in general. 

The open-source TempestExtremes tracking algorithm \citep{Ullrich_Zarzycki_2017} is used to detect and track individual extratropical cyclones using similar criteria described in \citet{zarzycki2018}. Candidate cyclones are determined by searching for minima in sea level pressure with a closed contour of 2 hPa within 6 great circle degrees of the minimum. Candidate cyclones, which are detected at 3-hourly increments, are then stitched together in time by searching within an 8-degree great circle radius at the next time increment for another candidate cyclone to form a cyclone track. For a cyclone track to be included in the analysis it must exist for at least 9 time slices representing a minimum cyclone track length of 24 hours.

EKE at 850 hPa is calculated by 
\begin{equation} \label{eke}{EKE = \frac{1}{2}(u\;'^2+v\;'^2)},\end{equation} 
where $(u\;',\; v\;')=(u-\overline{u},\; v-\overline{v})$ represent zonal and meridional velocity deviations from the annual mean velocities $(\overline{u},\; \overline{v})$ that obtained by averaging the velocities $(u,\; v)$ over the entire study period (1980--2014). Calculating $(u\;',\; v\;')$ with respect to a multi-year, or yearly, or moving seasonal average does not change the amplitude or phase of the seasonal cycle of EKE significantly \citep{Rieck_2015}. Following past work \citep{blackmon_1976, Ulbrich_2008,  Harvey_2014, Schemm_2018}, we apply a 2--6-day Butterworth bandpass filter \citep{russell_2006} to $(u\;',\; v\;')$ to retain reasonable timescales of extratropical cyclone activities. Other bandpass ranges, such as 2--8-day \citep{Yin_2005} and 3--10-day \citep{Kaspi_2013, Tamarin_2016}, are also tested. EKE is quantitatively sensitive to the bandpass range: longer range (e.g., 2--8-day vs. 2--6-day) translates to larger EKE, but the qualitative spatial pattern and seasonal variation of EKE are not sensitive to these ranges (not shown).

\subsubsection{Synoptic Composites for Extreme Cases}

We calculate the composite of synoptic anomalies conditioned on extreme SLS environments within six 5$^\circ$$\times$$5^\circ$ regions (i.e., R1--R6 in Figure \ref{fig_map}b) from ERA5 and CAM6. R2--R4 are selected following \citet{Ribeiro_Bosart_2018}; we also select sub-regions over the northern Great Plains (R1), the Midwest (R5), and the southeastern United States (R6), so that our analysis spans much of the land east of the Rocky Mountains where SLS activity and environments are concentrated. In this study, we define an extreme case in a region when the CAPES06 exceeds its local 99th percentile (i.e., within the top 1\%) in at least 80\% of the total grid points within the region. To generate the composite synoptic anomalies for each region, we first calculate the synoptic anomalies of each case from the full-period (1980--2014) monthly mean state (e.g., for a case selected in July 2012, the anomaly field of a variable is the difference between the variable field from the case and the mean field of the variable during 1980-2014 in July). Then, we generate composite synoptic anomalies at 250 hPa (horizontal winds and geopotential height), 700 hPa (horizontal winds, temperature, and geopotential height), and surface (10-m winds, 2-m specific humidity, and sea level pressure) by averaging the anomaly fields of the extreme cases for each region. The composite synoptic patterns based on a 50\% threshold is qualitatively similar, though with less distinctive features (e.g., a smoothed trough or wind fields; not shown). The 50\% threshold produces more candidates for each region than the 80\% threshold does, but also introduces larger variance that may reduce similarity across cases.

\section{Results: Radiosonde and ERA5}\label{sec:results_era5} 

\begin{figure}[t]
\centerline{\includegraphics[width=19pc]{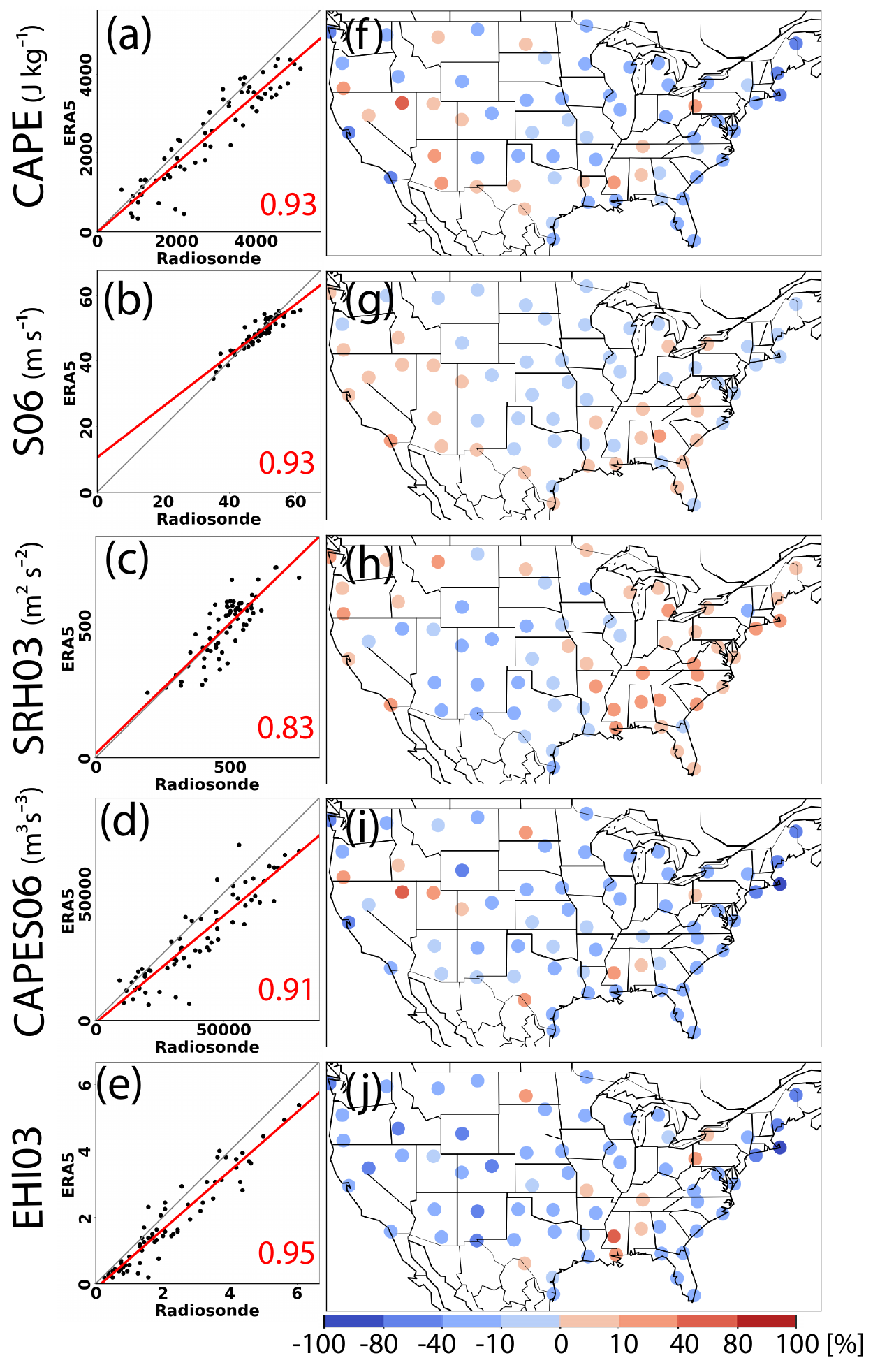}}
\caption{ERA5 reanalysis vs. radiosonde observations for the 99th percentiles of (top to bottom) CAPE, S06, SRH03, CAPES06, and EHI03. (a--e) Scatter plot (black dots) of the 99th-percentile values from ERA5 and radiosondes at each site (66 sites in total) with linear least-squares fit (red line) and pattern correlation coefficient (red text); gray line denotes a one-to-one fit. (f--j) Bias of the ERA5 99th percentile for each site, defined as percentage difference (ERA5 minus radiosonde). The 99th percentiles are generated from 0000 and 1200 UTC data during 1998--2014 for each site. Sample size from radiosondes at each site is shown in supplementary Fig. S1a. ERA5 99th percentiles are first generated at ERA5 grids and then linearly interpolated onto radiosonde sites.}
\label{fig_radiosonde_99ile}
\end{figure}

We begin by comparing the extreme values (99th percentile) of SLS environmental parameters and proxies in the ERA5 reanalysis against radiosondes to investigate the extent to which the ERA5 reanalysis can reproduce the observed SLS environments. We first calculate the pattern correlation coefficient between ERA5 and radiosondes, as well as the bias (defined as percentage difference from radiosonde value) in ERA5 for each parameter and proxy (Figure \ref{fig_radiosonde_99ile}). ERA5 in general performs well in reproducing extreme values of these constituent parameters, with strong pattern correlation (CAPE: 0.93; S06: 0.93; SRH03: 0.83; Figure \ref{fig_radiosonde_99ile}a--c) and relatively low bias, especially over the Great Plains ($\sim$ {$\pm${10\%}}; Figure \ref{fig_radiosonde_99ile}f--h). Specifically, extreme CAPE is generally underestimated (-10\%--0 for the central Great Plains; -40\%--\;-10\% for other areas) for most stations east of the Rocky Mountains and overestimated ($\sim$ {40\%}) over high terrains to the west (Figure \ref{fig_radiosonde_99ile}f); extreme S06 has the smallest bias (within $\pm${10\%}) across most stations (Figure \ref{fig_radiosonde_99ile}g); extreme SRH03 is generally underestimated over central and western United States with relatively low bias over the Great Plains (within $\pm${10\%}), while overestimated over eastern United States (10\%--40\%; Figure \ref{fig_radiosonde_99ile}h). Owing to the strong pattern correlations in these constituent parameters, the combined proxies, CAPES06 and EHI03, also have strong pattern correlations (CAPES06: 0.91; EHI03: 0.95), though they are in general underestimated by ERA5 (Figure \ref{fig_radiosonde_99ile}d, e). Biases of extreme CAPES06 is similar to the biases of extreme CAPE; biases of extreme EHI03 are slightly enhanced due to the combined influence of the constituent parameters, with negative bias of generally -40\% --\;-10\% across most stations over the central United States and -80\% --\;-40\% over western and northeastern United States (Figure \ref{fig_radiosonde_99ile}i, j). 

\begin{figure}[t]
\centerline{\includegraphics[width=19pc]{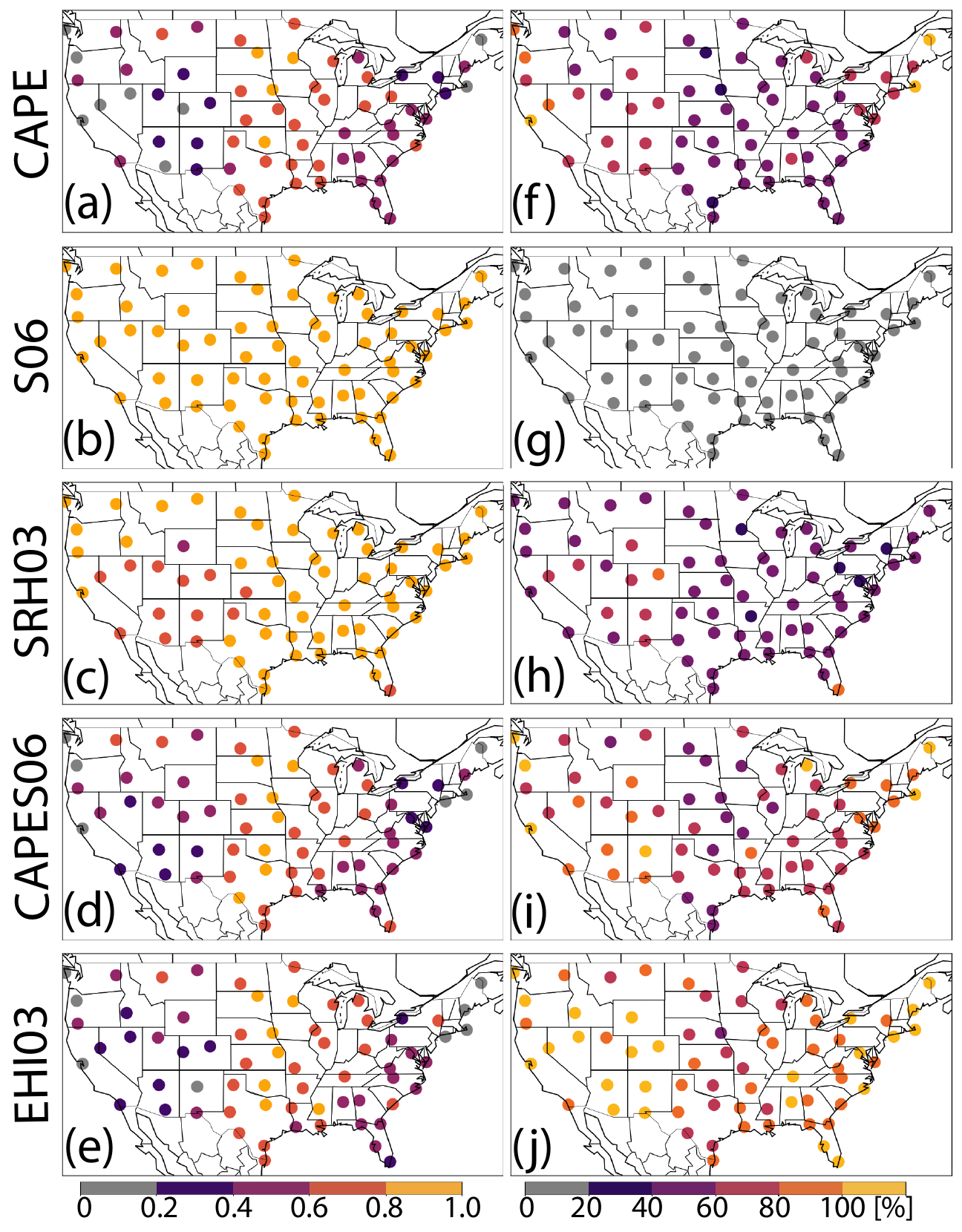}}
\caption{ERA5 reanalysis vs. radiosonde observations for the full-period (1998--2014) time series of (top to bottom) CAPE, S06, SRH03, CAPES06, and EHI03. (a--e) Temporal correlation coefficient. (f--j) Root Mean Square Error (RMSE) normalized by the local temporal mean radiosonde value. For CAPE (and thus CAPES06 and EHI03), cases with CAPE $\geq$ 500 J kg$^{-1}$ from radiosondes are evaluated (sample size: supplementary Fig. S1c); S06 and SRH03 evaluation is given for all cases after quality control (sample size: supplementary Fig. S1a). ERA5 time series are generated by linearly interpolating ERA5 values onto radiosonde sites at each time step (0000 and 1200 UTC) where qualified radiosonde values exist.}
\label{fig_radiosonde_overall}
\end{figure}

Similar good performance is found for the full-period (1998--2014) time series of each parameter and proxy at each station in terms of the temporal correlation (Figure \ref{fig_radiosonde_overall}a--e) and root-mean-square error (RMSE; Figure \ref{fig_radiosonde_overall}f--j). Here we condition the radiosondes on CAPE $\geq$ 500 J kg$^{-1}$, as our focus is on the environments that could support SLS events. This threshold is only applied to CAPE (and thus CAPES06 and SRH03), resulting in a smaller sample size at each site for these parameters (Figure S1c) than for S06 and SRH03 (Figure S1a). ERA5 has relatively strong temporal correlations ($\sim$ {0.8}) and small RMSEs ($\sim$ {40\%}) for CAPE over the Great Plains and the Midwest as compared to other areas (Figure \ref{fig_radiosonde_overall}a,f), and does an excellent job in representing S06 at all sites (temporal correlation $>$ 0.8; RMSE $<$ 20\%; Figure \ref{fig_radiosonde_overall}b,g) and SRH03 over eastern half of the United States (temporal correlation $>$ 0.8; 40\% $<$ RMSE $<$ 60\%; Figure \ref{fig_radiosonde_overall}c,h). Regarding CAPES06 (Figure \ref{fig_radiosonde_overall}d,i) and EHI03 (Figure \ref{fig_radiosonde_overall}e,j), stations over the Great Plains in general have strong temporal correlation ($>$0.8) and relatively low RMSE (CAPES06: 40\%--60\%; EHI03: 60\%--80\% ) as compared to the western United States and eastern coastal areas. Additionally, we perform similar analyses but condition CAPE, CAPES06, and EHI03 on CAPE $\geq$ 0 J kg$^{-1}$, as SLS events also occur within low-CAPE environments (e.g., the SLS activity commonly associated with low-CAPE, high-shear environments over the Southeast). By including these low-CAPE cases, both the temporal correlations and RMSEs of CAPE, CAPES06, and EHI03 increase (Figure S2), indicating persistent and relatively large biases for the low-CAPE cases. Note though that including cases with CAPE below 500 J kg$^{-1}$ skews the majority of our dataset to these relatively low CAPE values and so is much less representative of significant SLS environments over Great Plains.  

Overall, ERA5 performs reasonably well in reproducing the spatiotemporal variability of key SLS environmental parameters and proxies when compared against radiosonde data, particularly east of the Rocky Mountains where SLS activity is most common. Kinematic parameters (S06 and SRH03) are generally better estimated than thermodynamic parameters (CAPE) by the ERA5 reanalysis, particularly in the mountain west and eastern coasts, due to resolution limitations and associated intrinsic difficulties representing thermodynamic variability in complex terrains \citep{Taszarek_etal_2018}. This is similar to the performance of other reanalyses \citep{Gensini_etal_2014, Taszarek_etal_2018}. The temporal analyses (Figure \ref{fig_radiosonde_overall}) indicate that there may be significant errors in the values of SLS environmental parameters and proxies at any given point in time at a given location. However, for climatological studies such as this work, it is the high percentiles (Figure \ref{fig_radiosonde_99ile}) that are most important to adequately represent.

\section{Results: ERA5 and CAM6}\label{sec:results_ctrl}

\begin{figure}[t]
\centerline{\includegraphics[width=19pc]{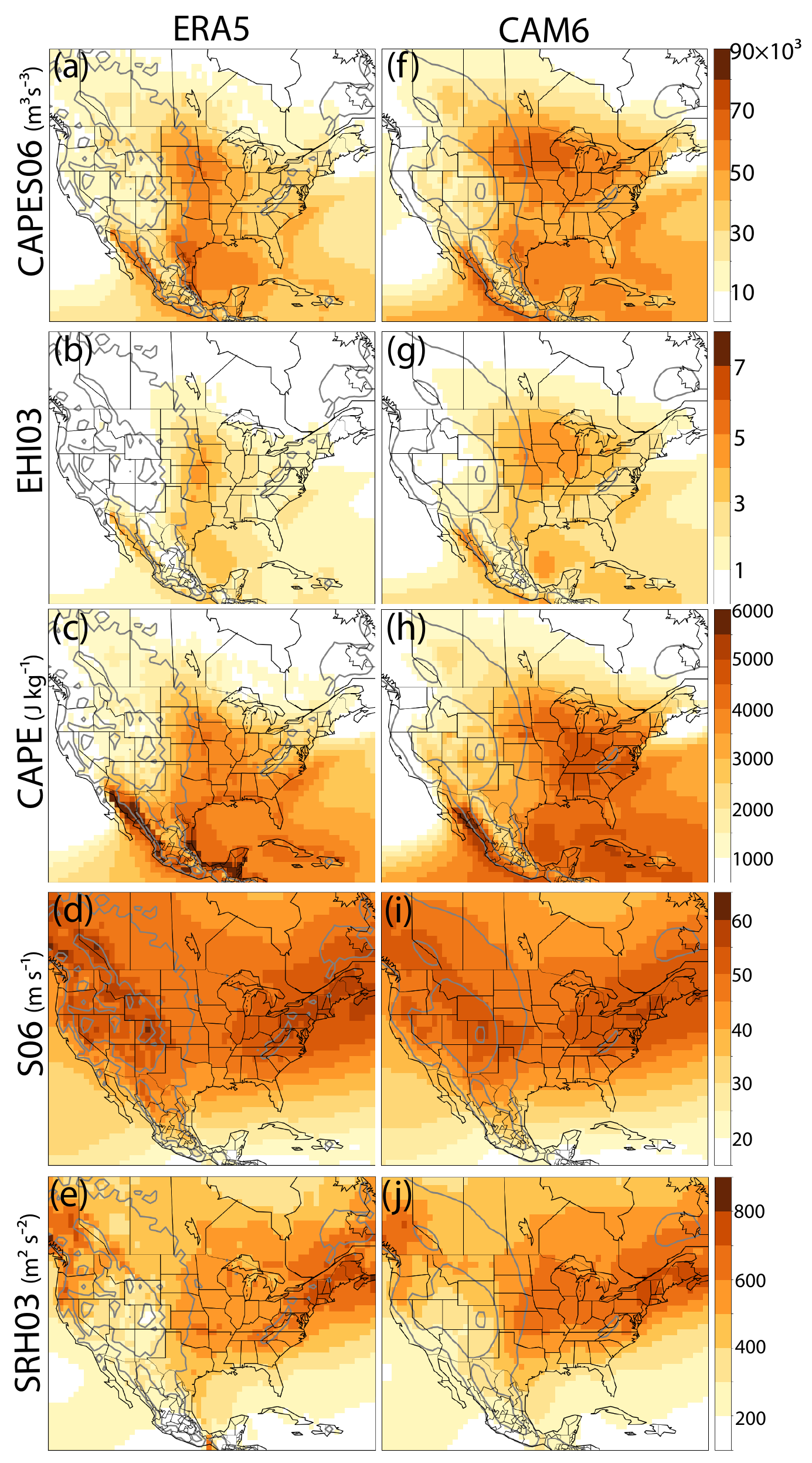}}
\caption{ERA5 reanalysis vs. CAM6 simulation for the 99th percentiles of (top to bottom) CAPES06, EHI03, CAPE, S06, and SRH03. (a--e) for ERA5, (f--j) for CAM6. 99th percentiles are generated at each grid point from the 3-hourly full-period (1980--2014) time series. Gray contour lines denote elevations at 500, 1500, and 2500 m. Differences (CAM6 minus ERA5) are shown in supplementary Fig. S3. }
\label{fig_era5_99ile_annual}
\end{figure}

\begin{figure*}[t]
\centering
\centerline{\includegraphics[width=28pc]{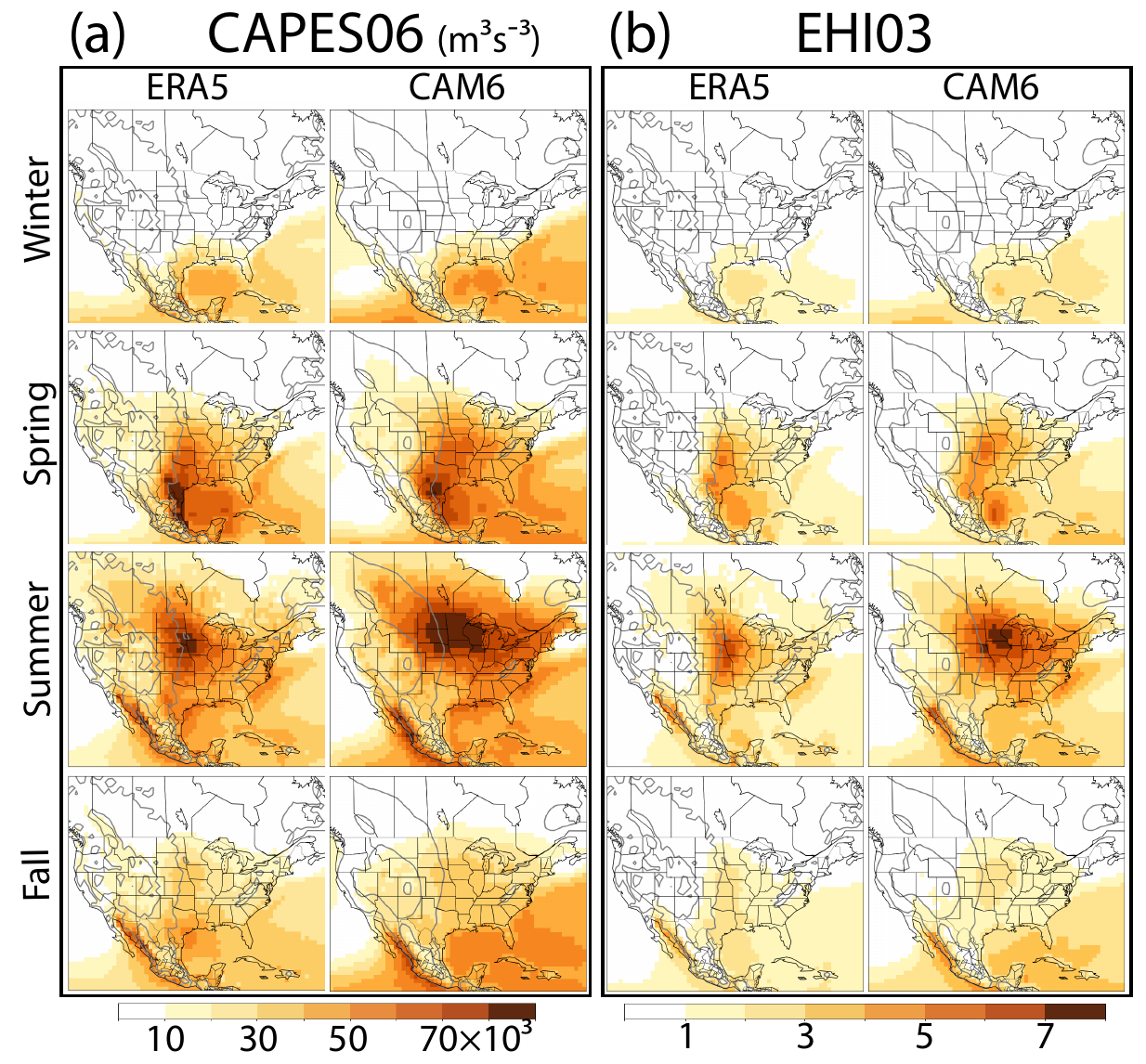}}
\caption{ERA5 reanalysis vs. CAM6 simulation for the seasonal 99th percentiles of (a) CAPES06 and (b) EHI03. Top to bottom: winter (DJF), spring (MAM), summer (JJA), and fall (SON). 99th percentiles are generated at each grid point from the 3-hourly full-period (1980--2014) time series during each season. Gray contour lines denote elevations at 500, 1500, and 2500 m. Differences (CAM6 minus ERA5) are shown in supplementary Fig. S6a--b.}
\label{fig_era5_seasonal_combined}
\end{figure*}

\begin{figure*}[t]
\centerline{\includegraphics[width=38pc]{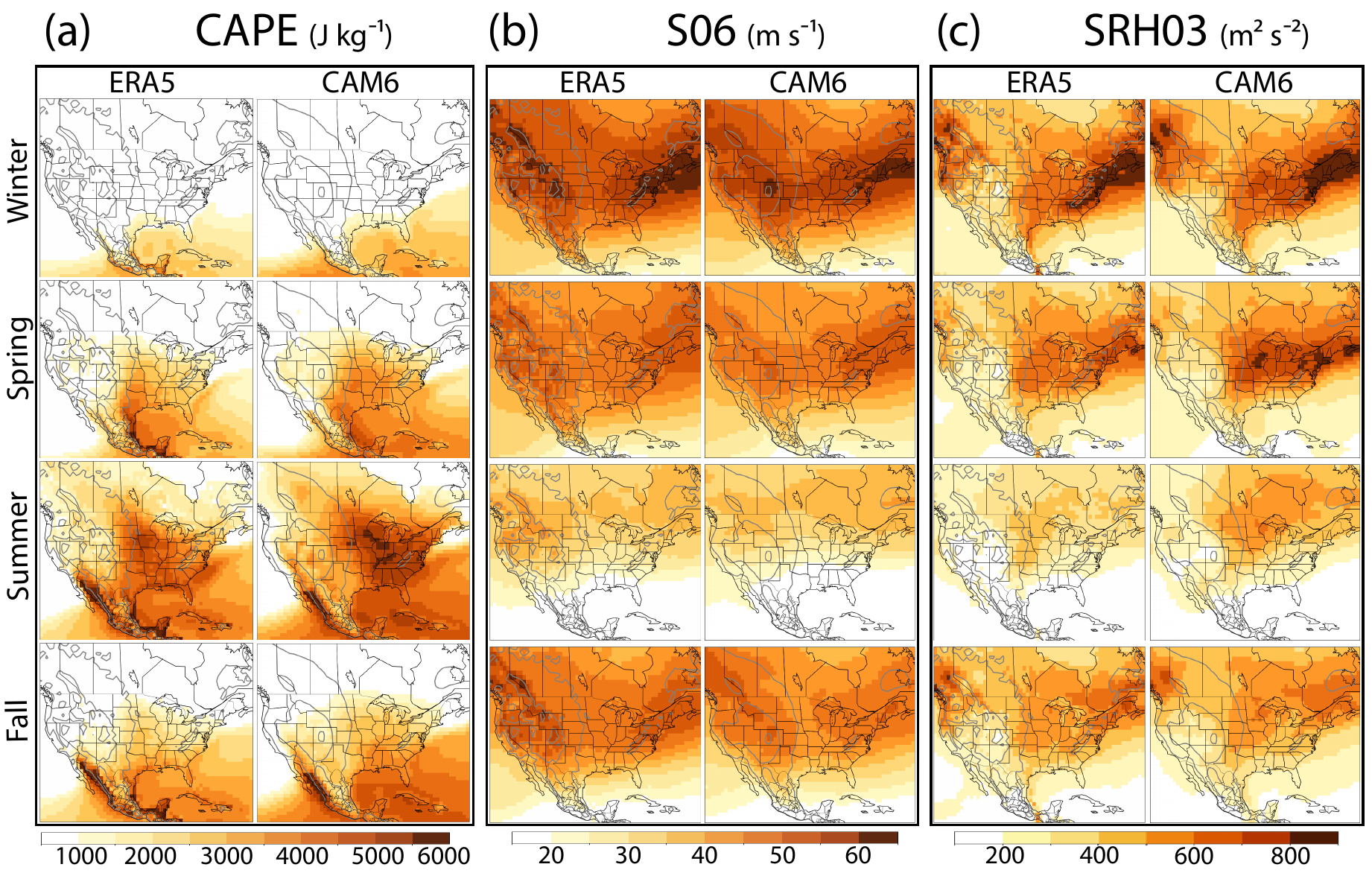}}
\caption{As in Fig. \ref{fig_era5_seasonal_combined}, but for (a) CAPE, (b) S06, and (c) SRH03. Differences (CAM6 minus ERA5) are shown in supplementary Fig. S6c--e.}
\label{fig_era5_seasonal_component}
\end{figure*}

\begin{figure*}[t]
\centerline{\includegraphics[width=24pc]{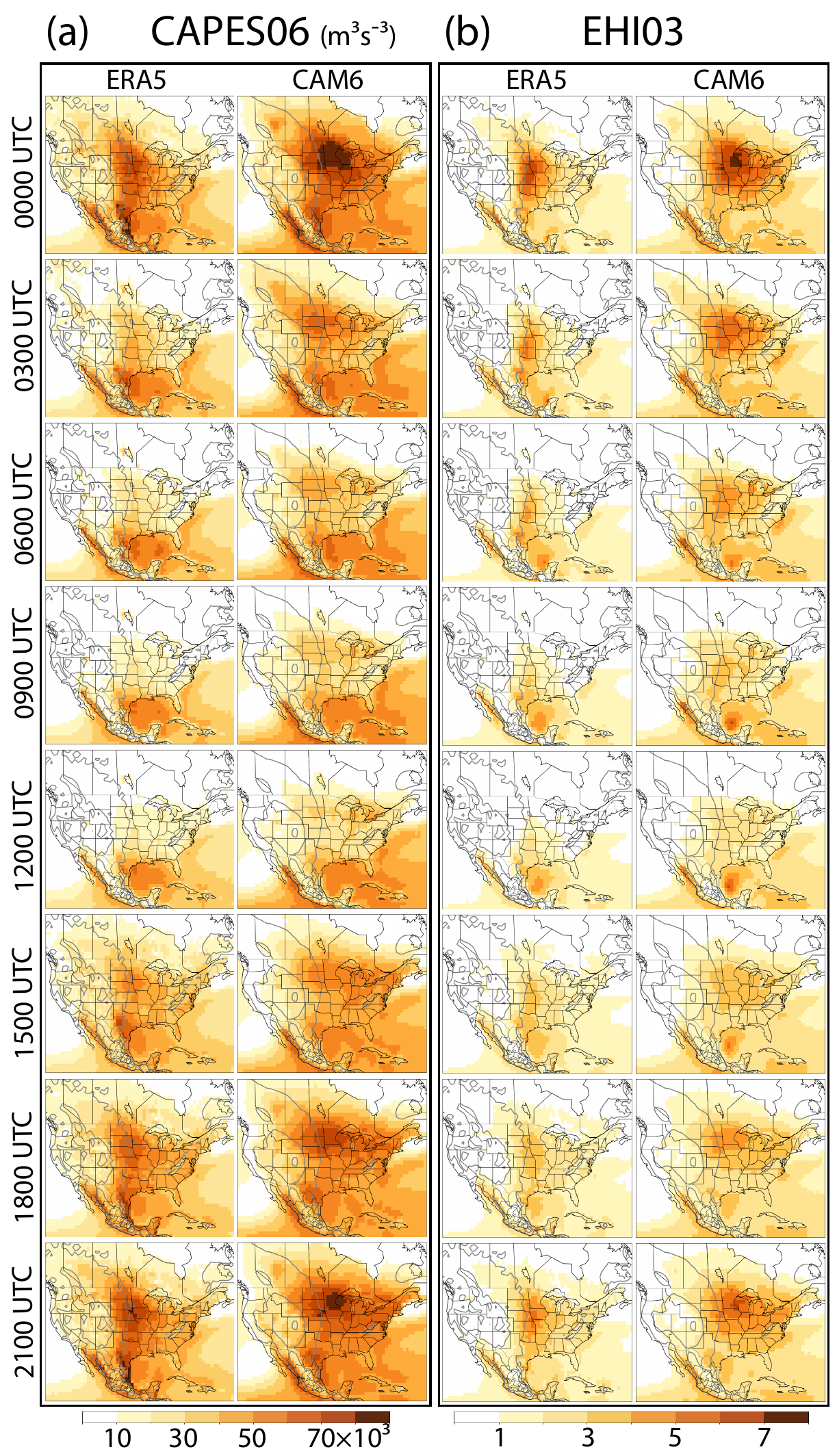}}
\caption{ERA5 reanalysis vs. CAM6 simulation for the diurnal 99th percentiles of (a) CAPES06 and (b) EHI03. Top to bottom: 0000, 0300, 0600, 0900, 1200, 1500, 1800, and 2100 UTC. 99th percentiles are generated at each grid point from full-period (1980--2014) time series at each 3-hourly UTC. Gray contour lines denote elevations at 500, 1500, and 2500 m. Differences (CAM6 minus ERA5) are shown in supplementary Fig. S8a--b.}
\label{fig_era5_diurnal_combined}
\end{figure*}

\begin{figure*}[t]
\centerline{\includegraphics[width=36pc]{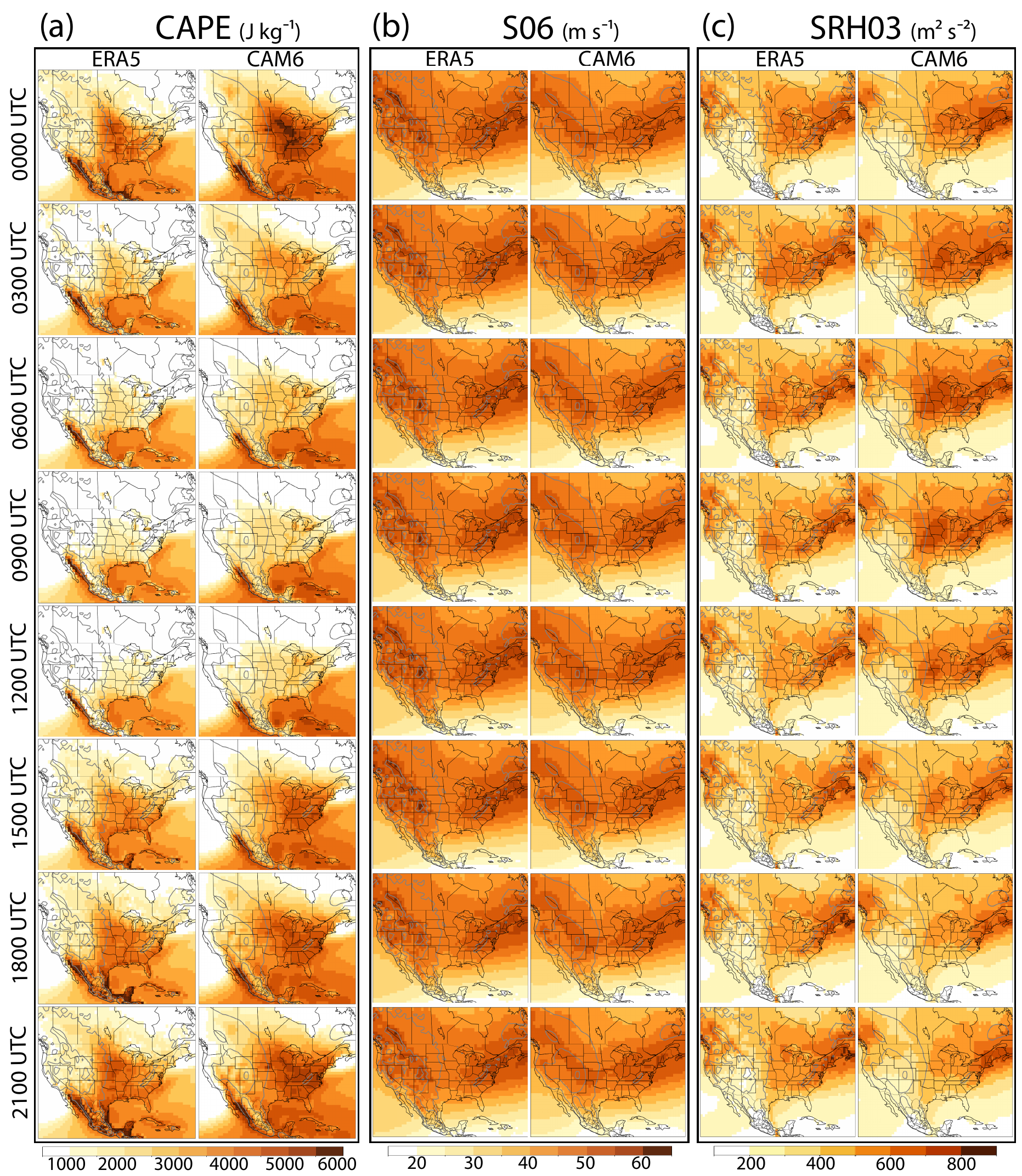}}
\caption{As in Fig. \ref{fig_era5_diurnal_combined}, but for (a) CAPE, (b) S06, and (c) SRH03. Differences (CAM6 minus ERA5) are shown in supplementary Fig. S8c--e.}
\label{fig_era5_diurnal_component}
\end{figure*}

\begin{figure*}[ht]
\centerline{\includegraphics[width=32pc]{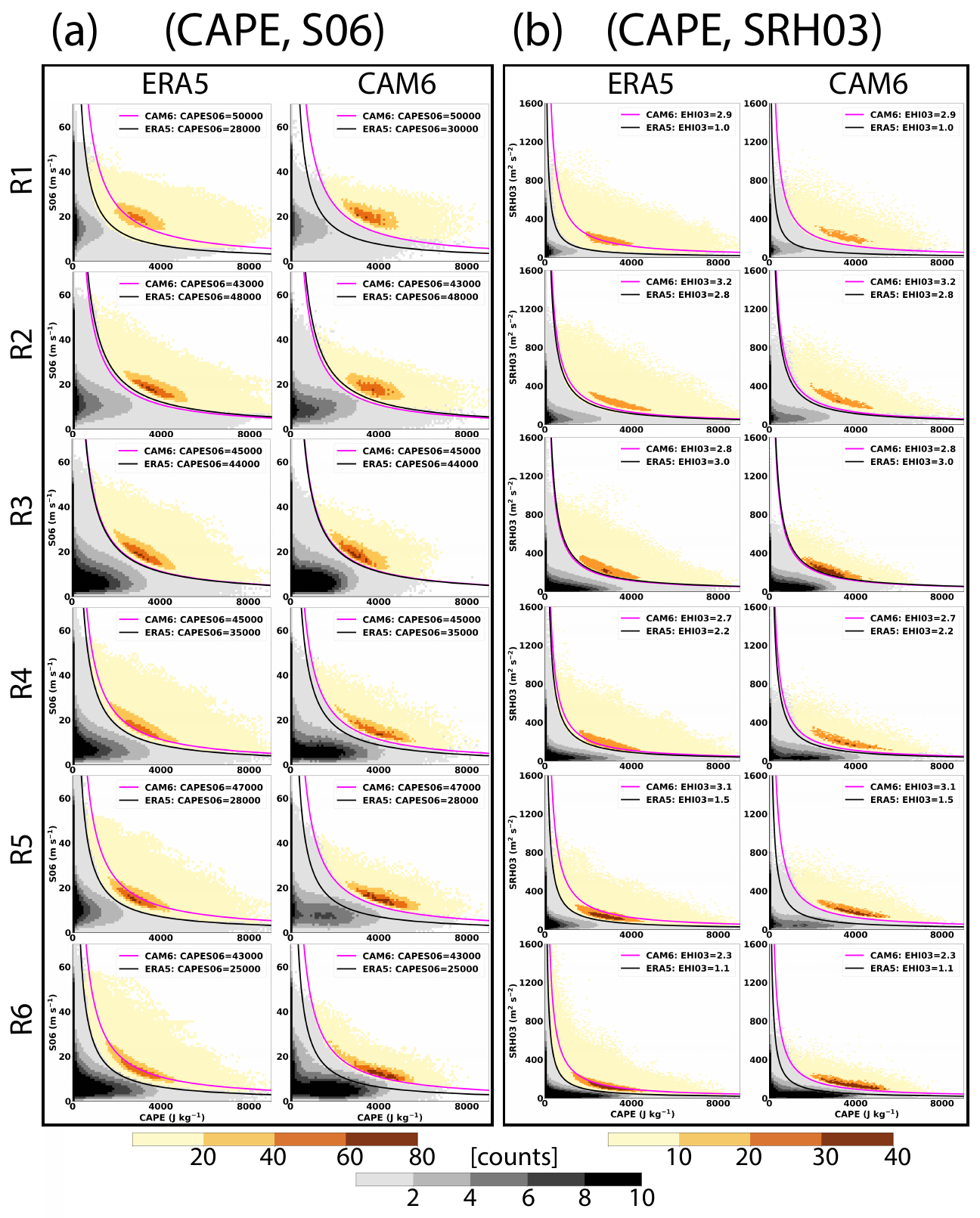}}
\caption{ERA5 reanalysis vs. CAM6 simulation for the joint probability distribution function (PDF; normalized to counts per 10,000 counts) of (a) (CAPE, S06) and (b) (CAPE, SRH03) for the top 1\% cases (colors) and all cases (grays) of CAPES06 and EHI03 from 1980--2014, respectively, in each of the sub-regions R1--R6 (as defined in Fig. \ref{fig_map}b). Solid lines represent the lower boundary of the top 1\% cases (i.e., the 99th percentile) of CAPES06 or EHI03 in ERA5 reanalysis (black) and CAM6 simulation (magenta).}
\label{fig_era5_jointpdf}
\end{figure*}

We now move on to an in-depth analysis of significant SLS environments and the associated synoptic-scale features over North America in both ERA5 reanalysis dataset and CAM6 simulation for period 1980--2014. Since the ERA5 reasonably reproduces SLS environments, based on its strong spatiotemporal correlations and relatively small biases with respect to radiosonde observations, especially over much of the eastern half of the United States where SLS activity is most prominent, this allows for an assessment of how CAM6 reproduces these climatological environments and the associated synoptic-scale features over North America as compared to the ERA5. Biases in CAM6 simulation with respect to the ERA5 are analyzed in terms of biases in the mean-state atmosphere. Additionally, common synoptic patterns that favor extreme SLS environments in regions east of the Rocky Mountains are analyzed.

\subsection{SLS Environments}

We first analyze the climatology of extreme values (99th percentile) of SLS environmental proxies and parameters over North America in ERA5 and CAM6. The spatial distributions of extreme CAPES06 and EHI03 in ERA5 indicate a similar climatological pattern of such environments (Figure \ref{fig_era5_99ile_annual}a, b). Both extreme CAPES06 and extreme EHI03 achieve a local maximum over southern Texas and over the central United States, consistent with that in the NCEP/NCAR reanalysis \citep{Brooks_etal_2003} and radiosondes \citep{Seeley_Romps_2015}. CAM6 simulation broadly reproduces the spatial pattern and amplitude of extreme CAPES06 and EHI03 in ERA5 (Figure \ref{fig_era5_99ile_annual}f, g). However, the local maximum extends farther east covering a larger area of eastern North America, with an enhancement of extreme CAPES06 and EHI03 primarily over the Upper Midwest (Figure \ref{fig_era5_99ile_annual}f, g and S3a, b).

CAM6 biases in extreme CAPES06 are predominantly tied to biases in extreme CAPE rather than S06, as CAM6 overestimates extreme CAPE (Figure \ref{fig_era5_99ile_annual}c, h and S3c) but does an excellent job in reproducing extreme S06 from ERA5 (Figure \ref{fig_era5_99ile_annual}d, i and S3d). Meanwhile, CAM6 biases in extreme EHI03 may be a result of biases in both extreme CAPE and SRH03, as CAM6 overestimates extreme SRH03 as well (Figure \ref{fig_era5_99ile_annual}e, j and S3e). Specifically, CAM6 simulates higher extreme CAPE over much of the eastern half of the United States than ERA5 does (Figure \ref{fig_era5_99ile_annual}c, h and S3c), and higher extreme SRH03 over the central United States and the Midwest (Figure \ref{fig_era5_99ile_annual}e, j and S3e); the simulated extreme S06 is nearly identical to ERA5, which attains its peak in a predominantly zonal band cutting through the central United States associated with the jet stream (Figure \ref{fig_era5_99ile_annual}d, i and S3d). We note that such spatial patterns of the proxies and parameters, as well as biases in CAM6 simulation with respect to ERA5 reanalysis, persist across the high-percentile cases (the 95th, 90th, and 75th percentiles), though the biases decrease moving toward lower percentiles (the 75th percentile; Figure S4). 

Though not our focus in this work, convective inhibition (CIN; defined as the negative integral of buoyancy from surface to the level of free convection) is also a key parameter associated with SLS activity and environments, as CIN provides a measure of the lower tropospheric stability that serves as possible barriers to the initiation of conditional instability \citep{williams1993,chen2020}. CIN extremes distribute broadly similar to CAPES06 and EHI extremes, and are overestimated by CAM6 as well (Figure S5). Deeper analysis for CIN representations in reanalysis datasets or climate models and the associated biases is desirable for future work.

We next analyze the seasonal cycle. Extreme CAPES06 and EHI03 in both ERA5 and CAM6 exhibit a strong seasonal cycle that peaks in warm seasons (spring and summer) (Figure \ref{fig_era5_seasonal_combined}), consistent with \citet{tippett_2015}. Specifically, the local maximum in spring occurs over southern Texas and shifts to the central United States in summer. A similar seasonal cycle is found in extreme CAPE (Figure \ref{fig_era5_seasonal_component}a), whereas the extreme S06 and SRH03 show an opposite seasonal phase that peaks in winter and reaches a minimum in summer (Figure \ref{fig_era5_seasonal_component}b, c). Biases in these SLS environments from CAM6 also exhibit a significant seasonal variation, as all proxies and parameters are biased higher in summer than in other seasons (Figure S6).  

Note that past work has also analyzed the number of days with significant SLS environments ($NDSEV$) to quantify SLS environments, such as the $NDSEV$ with CAPES06 $\geq$ 10,000 m$^{3}$ s$^{-3}$ \citep{Brooks_etal_2003, Trapp_etal_2007, Trapp_etal_2009, Diffenbaugh_etal_2013, Seeley_Romps_2015, Hoogewind_etal_2017} or the $NDSEV$ with CAPES06$\geq$ 20,000 m$^{3}$ s$^{-3}$ \citep{gensini_ashley_2011, gensini_ramseyer_2014,Gensini_Mote_2015}. The results of this method can differ from the high-percentile method used here \citep{Singh_etal_2017}, as $NDSEV$ is a pure frequency that does not account for the magnitude above threshold. For comparison, we calculate the climatological and seasonal cycle of $NDSEV$ with CAPES06 exceeding 10,000 m$^{3}$ s$^{-3}$ and 20,000 m$^{3}$ s$^{-3}$ respectively in ERA5 reanalysis (Figure S7). Relative to the 99th percentile, the annual $NDSEV$ with CAPES06 exceeding 10,000 m$^{3}$ s$^{-3}$ is shifted toward the southern United States with the local maximum confined to southern Texas (Figure S7a), whereas distributions of the $NDSEV$ with CAPES06 exceeding 20,000 m$^{3}$ s$^{-3}$ are broadly similar to distributions of the 99th percentile of CAPES06 (Figure S7f--j). This is likely because cases with the lower threshold of CAPES06 are weighted more by CAPE than S06, and thus the distribution of $NDSEV$ is dominated by that of high CAPE which both show small seasonal variations over southern Texas, especially in spring and summer (Figure S7b--e and 7a). The $NDSEV$ with EHI03 $\geq$ 1 is calculated as well, which indicates similar distribution pattern to EHI03 extremes (Figure S7k--o). 

We next analyze the diurnal cycle. Extreme CAPES06 (Figure \ref{fig_era5_diurnal_combined}a) and EHI03 (Figure \ref{fig_era5_diurnal_combined}b) in both ERA5 and CAM6 exhibit a strong diurnal cycle, particularly in the continental interior where the diurnal cycle peaks during the late afternoon to early evening (2100--0000 UTC) and reaches a minimum in the early morning (0900--1200 UTC). Such diurnal cycle behavior also exists along the Gulf and Atlantic coasts but with a much smaller amplitude, resembling the diurnal variation over ocean. These results are in line with past work on the diurnal variation of deep convection over North America \citep{wallace_1975, nesbitt_2003, Tian_etal_2005}. The diurnal cycle of extreme CAPES06 and EHI03 is dominated by that of extreme CAPE (Figure \ref{fig_era5_diurnal_component}a), whereas the amplitude of the diurnal cycle of extreme S06 (Figure \ref{fig_era5_diurnal_component}b) is relatively small and extreme SRH03 peaks later (at around 0600 UTC) than extreme CAPES06 and EHI03 (Figure \ref{fig_era5_diurnal_component}c). The diurnal cycle signal of extreme SRH03 is strongest over the central United States, associated with the diurnal oscillation of the Great Plains low-level jets. Biases in these SLS environments from CAM6 also exhibit diurnal variations similar to the behaviors of the proxies and parameters themselves (Figure S8).

How is extreme CAPES06 or EHI03 affected by its constituent parameters? To more precisely answer this question, we analyze the joint PDF of (CAPE, S06) and (CAPE, SRH03) within each box (R1--R6 in Figure \ref{fig_map}b). We analyze the entire joint distribution with special emphasis added to the top 1\% cases of CAPES06 and EHI03. Both ERA5 and CAM6 indicate that the CAPES06 and EHI03 extremes consist of large-to-extreme CAPE but moderate-to-small S06 and SRH03 (Figure \ref{fig_era5_jointpdf}), as was found by \citet{Diffenbaugh_etal_2013}. Meanwhile, the S06 and SRH03 extremes are associated with small values of CAPE, corresponding to the high-shear, low-CAPE environments (defined as CAPE $\leq$ 500 J kg $^{-1}$ and S06 $\geq$ 18 m s $^{-1}$) \citep{guyer_2010, sherburn_etal_2014, Sherburn_etal_2016}. These results are also evident in the seasonal cycles described above (Figure \ref{fig_era5_seasonal_combined}, \ref{fig_era5_seasonal_component}): high CAPES06 and EHI03 in summer are associated with very high CAPE but relatively low S06 and SRH03, while the high-shear, low-CAPE environments, corresponding to low CAPES06 and EHI03, are concentrated in winter over land. The joint PDFs also indicate that CAPES06 and EHI03 extremes are greater in CAM6 than in ERA5 for regions over the northern Great Plains (R1) and southeastern United States (R5, R6), where the joint PDF shifts toward higher (CAPE, S06) and (CAPE, SRH03). Meanwhile, the difference in the PDFs between ERA5 and CAM6 is relatively small over south-central United States (R2--R4).

\subsection{SLS-Relevant Synoptic-Scale Features} 

\begin{figure}[b]
\centerline{\includegraphics[width=18pc]{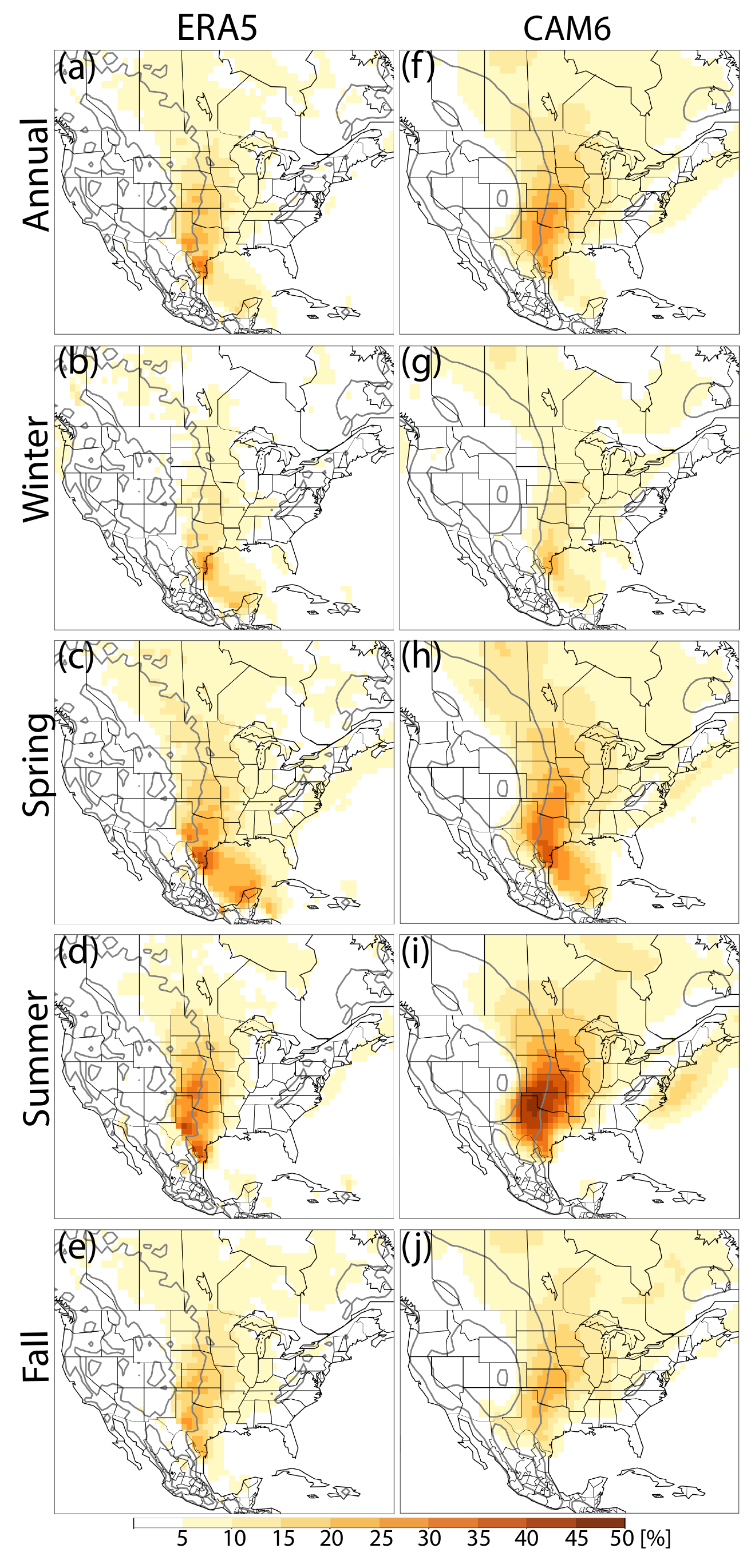}}
\caption{ERA5 reanalysis vs. CAM6 simulation for the mean frequency percentages of southerly Great Plains low-level jets during 1980--2014. For ERA5: (a) annually, (b) winter (DJF), (c) spring (MAM), (d) summer (JJA), and (e) fall (SON). (f-j) as in (a-e) but for CAM6. Gray contour lines represent elevations at 500, 1500, and 2500 m.}
\label{fig_era5_seasonal_gpllj}
\end{figure}

\begin{figure}[b]
\centerline{\includegraphics[width=18pc]{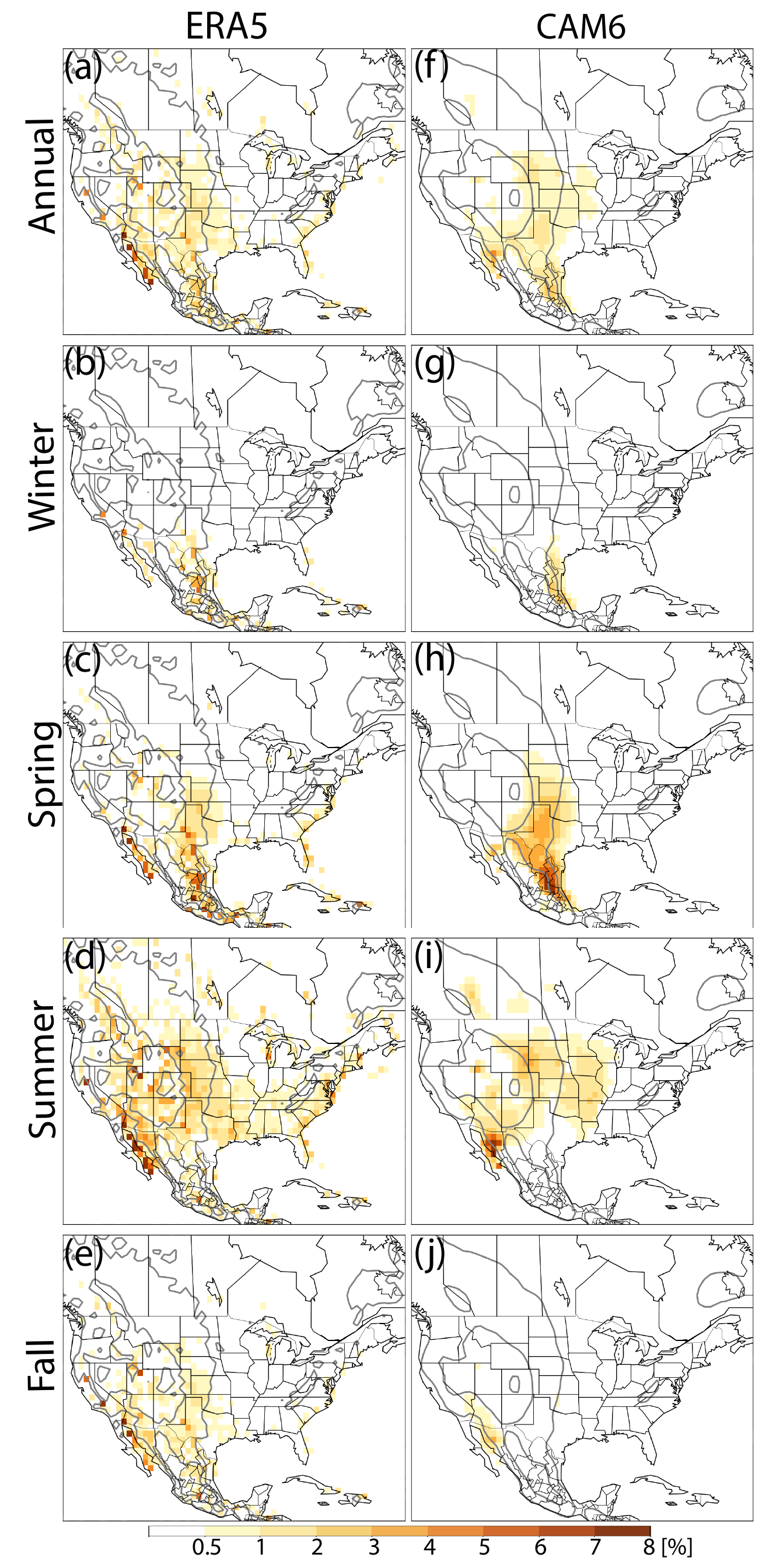}}
\caption{As in Fig. \ref{fig_era5_seasonal_gpllj}, but for drylines.}
\label{fig_era5_seasonal_dryline}
\end{figure}

\subsubsection{Southerly GPLLJ}

ERA5 southerly GPLLJ frequency is concentrated primarily over the central and southern Great Plains (Figure \ref{fig_era5_seasonal_gpllj}a), with a seasonal frequency peak (up to 40\%) in spring and summer (Figure \ref{fig_era5_seasonal_gpllj}b--e). The local maximum is located in southern Texas in spring and shifts to southwestern Texas and western Oklahoma in summer, consistent with results of NARR reanalysis \citep{Walters_etal_2014, Doubler_etal_2015} and observations \citep{Bonner_1968, Walters_etal_2014}. CAM6 broadly reproduces the spatial pattern and amplitude of the southerly GPLLJ frequency (Figure \ref{fig_era5_seasonal_gpllj}f--j), except for summer when the frequency percentage in CAM6 (up to 50\% over western Oklahoma; Figure \ref{fig_era5_seasonal_gpllj}i) is much higher than that in ERA5. This increased occurrence of southerly GPLLJ indicates stronger mean low-level winds in CAM6, which contribute to the positive biases in SRH03 over the central United States as well. Meanwhile, more moisture could be transported from the Gulf of Mexico \citep{Helfand_Schubert_1995, Higgins_etal_1997}, which may partially explain the enhanced CAPE in CAM6.   

\subsubsection{Dryline}

ERA5 dryline frequency is also concentrated in the central and southern Great Plains (Figure \ref{fig_era5_seasonal_dryline}a), with a seasonal frequency peak (up to 8\%) in spring over southwestern Texas (Figure \ref{fig_era5_seasonal_dryline}c). The distribution shifts poleward into the central plains in summer with a reduced frequency percentage (4\%) but spanning a larger area (Figure \ref{fig_era5_seasonal_dryline}d). These results are qualitatively similar to observations by \citet{Schaefer_1974} and \citet{Hoch_Markowski_2005}, though the amplitude of dryline occurrence is lower than \citet{Hoch_Markowski_2005} owing to the stricter criteria used in this work in an effort to distinguish drylines from fronts. CAM6 performs well in reproducing the dryline distribution, as the spatial pattern and amplitude of dryline frequency over the Great Plains and its seasonal variation are broadly similar to that in ERA5 (Figure \ref{fig_era5_seasonal_dryline}f--j). ERA5 appears to better identify the less-frequent drylines over the far eastern United States \citep{Duell_Broeke_2016}, perhaps owing to its higher horizontal resolution that may permit detection of smaller-scale gradients in surface specific humidity. Meanwhile, ERA5 identifies a number of strong horizontal moisture gradient along the east coast, especially in summer, while these may not correspond with typical drylines over the Great Plains.   

\begin{figure}[t]
\centerline{\includegraphics[width=18pc]{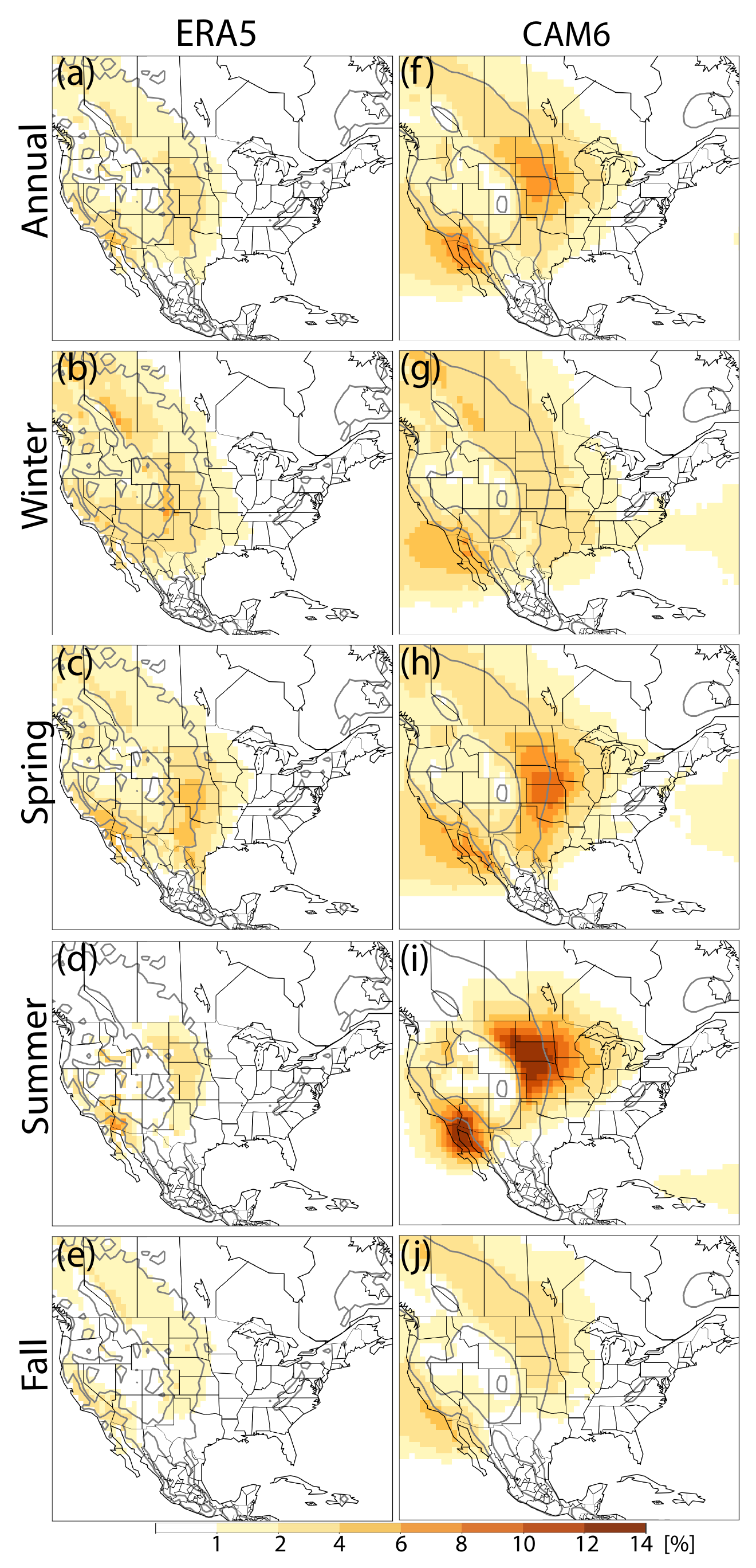}}
\caption{As in Fig. \ref{fig_era5_seasonal_gpllj}, but for elevated mixed layers.}
\label{fig_era5_seasonal_eml}
\end{figure}

\subsubsection{EML}

ERA5 EML frequency is again concentrated over the Great Plains (Figure \ref{fig_era5_seasonal_eml}a), with a seasonal frequency peak in spring over south-central United States (6\%; Figure \ref{fig_era5_seasonal_eml}c), consistent with \citet{Lanicci_Warner_1991a}. The local maximum shifts poleward to South Dakota and Nebraska in summer with a reduced frequency percentage (4\%; Figure \ref{fig_era5_seasonal_eml}d), qualitatively similar to findings of \citet{Ribeiro_Bosart_2018}. Note that the amplitude of EML occurrence is sensitive to the identification criteria: the criteria with 7.5 K km$^{-1}$ for the minimum lapse rate and 150 hPa for the minimum EML depth significantly increase the frequency percentage of EML in ERA5 (e.g., 24\% in spring and 20\% in summer; not shown). CAM6 successfully reproduces these spatial patterns (Figure \ref{fig_era5_seasonal_eml}f--j). However, it exhibits a significant positive bias, particularly in summer (14\% over South Dakota), which is associated with the generally larger mid-level lapse rate in CAM6 than in ERA5 (analyzed below in Figure \ref{fig_era5_seasonal_profile}). These enhanced EMLs potentially produce more CAPE, which may also partially explain the enhanced CAPE in CAM6.

\begin{figure}[t]
\centerline{\includegraphics[width=18pc]{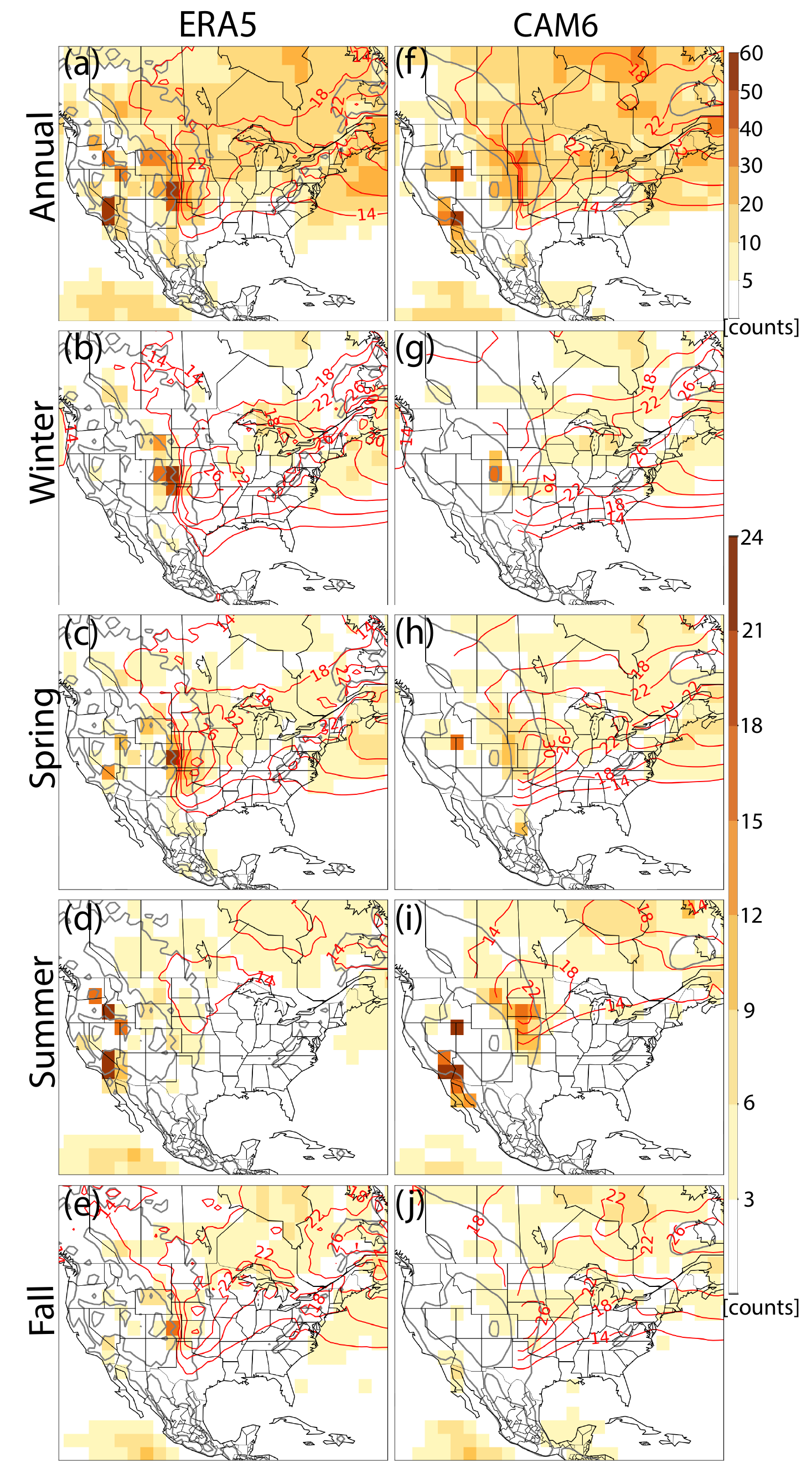}}
\caption{As in Fig. \ref{fig_era5_seasonal_gpllj}, but for cyclone tracks (counts per 2.5$^\circ$$\times$$2.5^\circ$ grid box; filled contours) and mean 2--6-day Butterworth bandpass filtered eddy kinetic energy at 850 hPa (m$^{2}$ s$^{-2}$, red contour lines).}
\label{fig_era5_seasonal_stormtrack}
\end{figure}

\subsubsection{Extratropical cyclone activity}

ERA5 cyclone track frequency has its primary local maximum to the lee of the Rocky Mountains in both the annual mean (Figure \ref{fig_era5_seasonal_stormtrack}a) and through the seasonal cycle (Figure \ref{fig_era5_seasonal_stormtrack}b--e), which is linked to cyclogenesis on the leeside of the Rocky Mountains. Secondary local maxima are also found over the Great Lakes and off the Northeastern United States coast, in line with findings of \citet{reitan_1974} and \citet{ zishka_1980}. Seasonal variation of cyclone tracks is characterized by a decrease in the frequency and a poleward shift of the local maximum from winter and spring (24 counts over southeastern Colorado) to summer (9 counts over Montana-South Dakota), qualitatively similar to past work \citep{reitan_1974, zishka_1980, Eichler_Higgins_2006}. Similar results are evident for EKE, as the core of EKE shifts from northwestern Oklahoma in winter to North Dakota in summer. CAM6 broadly reproduces the spatial pattern and amplitude of cyclone tracks and EKE (Figure \ref{fig_era5_seasonal_stormtrack}f--j). The cyclone track frequency over southeastern Colorado in CAM6 is smaller than that in ERA5 during winter, spring, and fall, likely due to the coarser horizontal resolutions \citep{chang_2013}. The major difference between CAM6 and ERA5 occurs in summer (Figure \ref{fig_era5_seasonal_stormtrack}d, i) when CAM6 produces more cyclone tracks and higher EKE over the central and northern Great Plains than ERA5 (roughly 21 vs. 9 counts and 22 vs. 14 m$^{2}$ s$^{-2}$), which may contribute to the positive bias in CAM6 SLS environments in summer relative to ERA5.  

\begin{figure*}[t]
\centerline{\includegraphics[width=38pc]{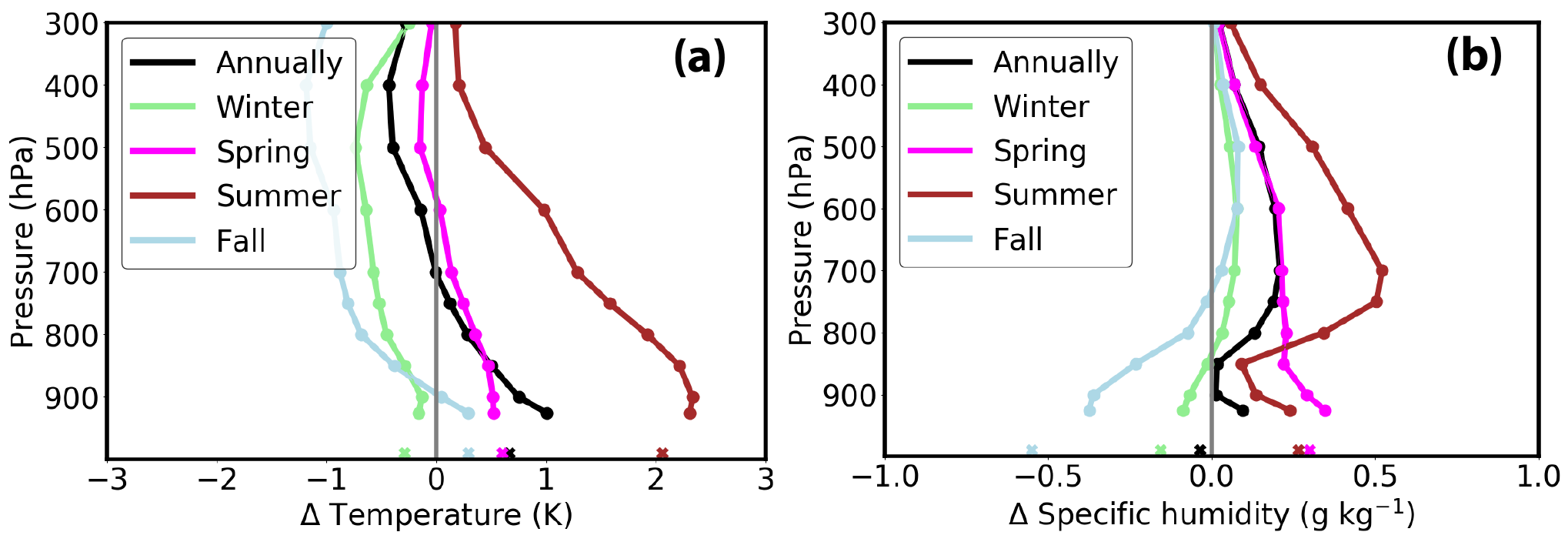}}
\caption{Difference (CAM6 minus ERA5) in the annual and seasonal mean profiles of (a) air temperature and (b) specific humidity over the eastern half of the United States (100--62$^\circ$W, 25--49$^\circ$N) during 1980--2014. Dots denote pressure levels of 925, 900, 850, 800, 750, 700, 600, 500, 400, and 300 hPa. Cross signs indicate the difference in 2-m temperature and specific humidity.}
\label{fig_era5_seasonal_profile}
\end{figure*}

Overall, both ERA5 and CAM6 produce qualitatively reasonable climatologies of the key synoptic-scale features over North America. The SLS environments are closely related to these synoptic-scale features as they have consistent climatological behavior in terms of their spatial distributions and seasonal cycles: (1) both SLS environments and the associated synoptic-scale features occur typically in warm seasons (spring or summer) to the east of the Rocky Mountains; and (2) they all exhibit a poleward shift of the local maximum from winter to summer and an equatorward shift again from summer to winter. Compared with ERA5, CAM6 overestimates SLS environments, principally due to a positive bias in extreme CAPE and to a lesser degree SRH03, particularly in summer over much of the eastern half of the United States. These results are consistent with the positive biases in most synoptic-scale features (GPLLJ, EML, and extratropical cyclone activity).   

\subsection{Biases in the mean-state atmosphere} 

In an effort to better understand these biases in CAM6, we analyze differences in the mean-state atmosphere, which may provide insight into underlying causes \citep{Trapp_etal_2007, Diffenbaugh_etal_2013}. Compared to ERA5, the CAM6 summer-mean atmosphere over the eastern half of the United States (100--62$^\circ$W, 25--49$^\circ$N; Figure \ref{fig_era5_seasonal_profile} and S9) is characterized by higher surface and lower-tropospheric temperatures (+2 K), enhanced low-to-mid-level (900--500-hPa) lapse rates (+0.5 K (100 hPa)$^{-1}$) that accounts for the overestimated EMLs, and higher specific humidity at the surface (+0.3 g kg$^{-1}$) and mid-levels (+0.4 g kg$^{-1}$ averaged through 800--400 hPa). The high biases in summertime extreme CAPE and the combined proxies in CAM6 is attributed primarily to increases in the surface specific humidity. This attribution is supported by high, statistically significant (p$<$0.001) linear pattern correlations ($r$) between biases in the extreme CAPE, CAPES06, and EHI03 and biases in the surface specific humidity over the eastern half of the United States ($r=$ 0.81, 0.72, and 0.80, respectively); generally small or statistically insignificant pattern correlations ($r<0.5$) are found with biases in surface temperature, low-to-mid-level lapse rate, or mid-level specific humidity. The summer-mean wind field at the surface and the upper levels in CAM6 and ERA5 is similar, though CAM6 produces slightly stronger upper-level jet streams over northern North America and slightly stronger surface onshore winds over Texas (Figure S9), helping explain the relatively small biases in extreme S06; while the southerly onshore winds over the Great Plains from the Gulf of Mexico are further enhanced in CAM6 at 900--850 hPa (not shown), consistent with the increased southerly GPLLJs and the positive bias of extreme SRH03 over the central Great Plains. Similar features are also evident in Spring though with smaller magnitude. As for winter and fall, the difference between ERA5 and CAM6 in the mean-state atmosphere is relatively small (Figure \ref{fig_era5_seasonal_profile} and S9), consistent with the comparable climatologies of SLS environments and synoptic-scale features analyzed above between ERA5 and CAM6 during these seasons.

These mean low-level warm and moist biases over the eastern half of the United States in CAM6 are associated with systematic warm and dry biases over the central United States while warm and moist biases over the eastern third of the United States (Figure S9a) that have been found to persist in many generations of regional and global climate models \citep{klein2006, cheruy2014, mueller2014, lin2017}. Explanation for such systematic bias over the central United States have been proposed, including soil moisture deficit \citep{koster2004, phillips2014} or precipitation deficit \citep{lin2017} that alters the lower-tropospheric mean state via land-atmosphere feedback processes in climate models \citep{koster2001, mo2003}. How precisely these systematic model biases over the eastern half of the United States affect the SLS environments and synoptic-scale features is a worthy topic left for future work.

\begin{figure*}[t]
\centerline{\includegraphics[width=36pc]{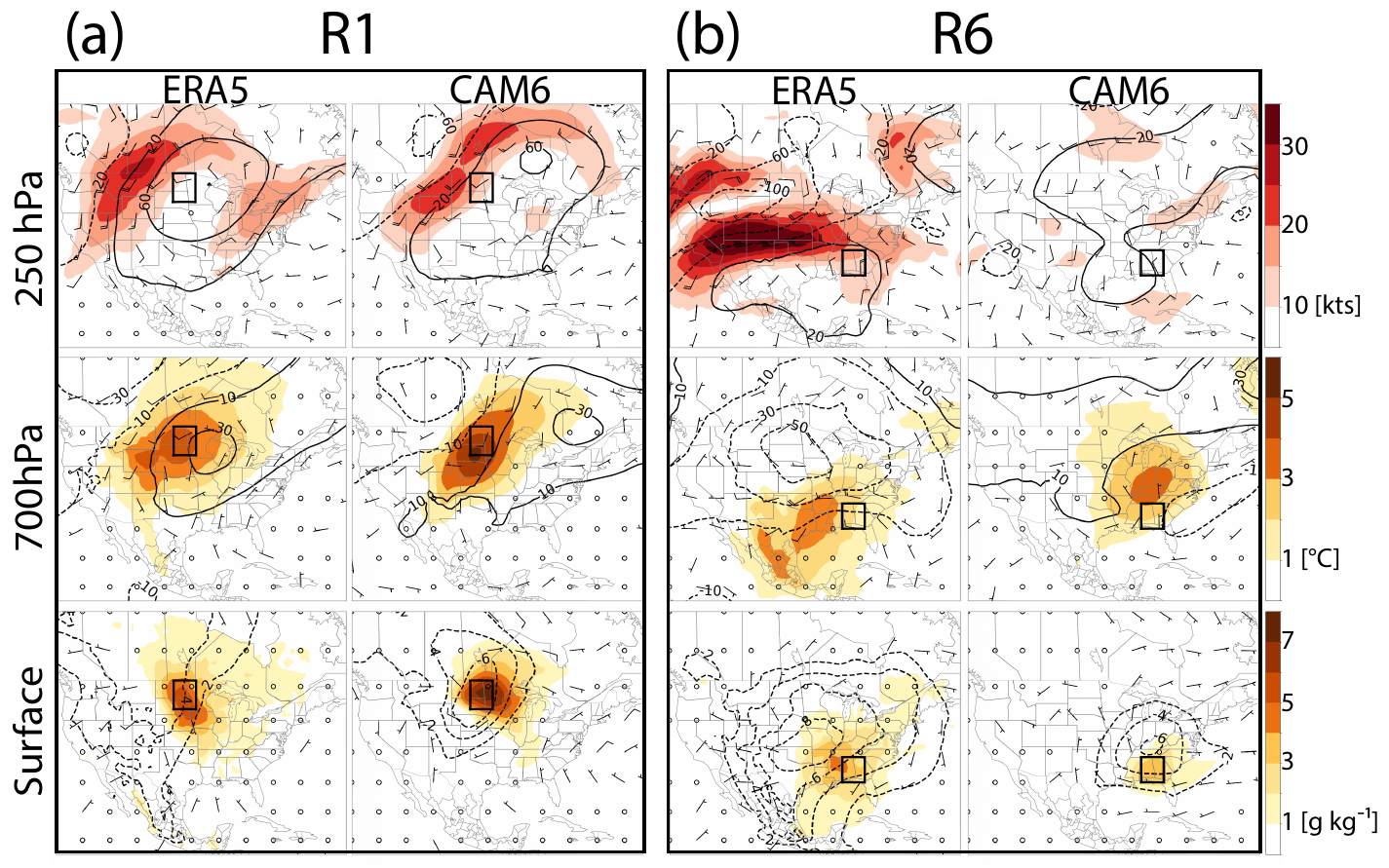}}
\caption{ERA5 reanalysis vs. CAM6 simulation for composite patterns of synoptic anomalies associated with significant SLS environments (details in the text) during 1980--2014 in regions (a) R1 and (b) R6. Black square shows the location of respective region (as defined in Fig. \ref{fig_map}b). Top row: 250 hPa, with composite anomalies of wind vector, wind speed (kts; filled contours), and geopotential height (m; black contour lines). Middle row: 700 hPa, with composite anomalies of wind vector, temperature ($^\circ$C; filled contours), and geopotential height (m; black contour lines). Bottom row: near surface, with composite anomalies of 10-m wind vector, 2-m specific humidity (g kg$^{-1}$; filled contours), and sea level pressure (hPa; black contour lines). Composite synoptic anomalies for sub-regions of R2--R5 are shown in supplementary Fig. S10.}
\label{fig_era5_80p}
\end{figure*}

\subsection{Synoptic Composites for Extreme Cases}

Finally, we compare composite patterns of synoptic anomalies associated with extreme SLS environments across our sub-regions over the eastern half of the United States (i.e., R1--R6 defined in Figure \ref{fig_map}b) between ERA5 and CAM6. Here, our analysis focuses on R1 and R6. R1 represents the region of the primary local maximum of SLS environments from the ERA5 reanalysis and CAM6 simulation, though the actual SLS occurrence maximum is further south \citep{Agee_etal_2016}; R6 is known to exhibit different behavior from the central United States and has shown a positive trend of SLS occurrence in recent decades \citep{Agee_etal_2016, Gensini_Brooks_2018}.

For region R1 over the eastern North Dakota and South Dakota (Figure \ref{fig_era5_80p}a; ERA5: 166 cases; CAM6: 92 cases), the ERA5 composite yields an enhanced ridge at 250 hPa whose axis extends from northern Texas to North Dakota, with an intensified jet streak along the United States-Canada border (+20--30-kts anomalies). This enhanced upper-level ridge forcing is associated with enhanced southwesterly flow at 700 hPa, which advects more warm and dry air from the elevated terrain eastward toward the Great Plains. Near the surface, the region is located on the southeast side of a trough anomaly extending southward from south-central Canada whose south-southwesterly flow has advected considerable moisture into the region. This composite pattern is similar to that found to be associated with progressive derechos \citep{johns1993,bentley2000,guastini2016}. CAM6 reproduces this composite pattern, though the 700-hPa warm advection is stronger and the surface trough anomaly is replaced by a slightly more intense cyclonic anomaly centered within the region. 

Composite synoptic patterns for cases in the central-southern Great Plains (R2--R4) and over Indiana (R5) are broadly similar between ERA5 and CAM6, and also indicate a similar setup to that in R1, though the relative position of synoptic features varies across regions (Figure S10). In general, a significantly intensified jet streak exists near the region at 250 hPa (except for R5 in CAM6 whose 250-hpa anomalies are relatively small). At 700 hPa, the region is located to the east or northeast of a warm air mass upstream, which supplies substantial warm air into the region due to the enhanced prevailing westerly or southwesterly winds. Near-surface air exhibits a deep trough or cyclonic anomaly with the region located to the east or southeast of the trough axis or the cyclonic center. As a result, robust advection of warm, moist air from the Gulf of Mexico into the region is found in the lower troposphere. These synoptic composites are broadly similar to the classic patterns of severe weather outbreaks over the Great Plains \citep{barnes1986,johns1992,johns1993,Mercer_etal_2012}.

The synoptic composite for sub-region R6 over the southeastern United States (Figure \ref{fig_era5_80p}b; ERA5: 40 cases; CAM6: 96 cases) differs from the other sub-regions as well as between ERA5 and CAM6. The ERA5 composite yields a much more intense jet stream (anomaly $\geq$ 30 kts) over the central United States at 250 hPa and a subtle shortwave trough at 700 hPa. Near the surface, a strong extratropical cyclone occupies much of the eastern half of North America with a low pressure anomaly centered over the central United States. The region (R6) is located to the southeast of the surface low. One key difference from the other regions is that this area is directly influenced by the flow on the west side of the North Atlantic Subtropical High, which enhances the southwesterly low-level winds and moisture advection over the region \citep{Stahle_Cleaveland_1992, Miller_mote_2017}. The CAM6 composite reproduces this near-surface flow pattern, but at higher levels it yields a broad ridge at 250- and 700-hPa with slight enhancement on the geopotential height and without any anomalous jet streak enhancement to the northwest of the region at 250 hPa. The surface cyclonic anomaly is weaker in CAM6 than in ERA5, causing less moisture transport into the region.

Overall, the common synoptic anomaly patterns associated with extreme SLS environments show small variations across sub-regions over much of the eastern half of the United States (R1--R5), where CAM6 also compares well with ERA5. The southeastern United States (R6) behavior differs somewhat from the other regions, as well as between ERA5 and CAM6, which implies differences in the generation of SLS environments in the Southeast as compared to farther inland. The SLS environments in the Southeast involve a significant portion of high-shear, low-CAPE environments \citep{guyer_2010, sherburn_etal_2014, Sherburn_etal_2016}, which are known to be associated with more difficult forecasting of SLS activity \citep{Miller_mote_2017}. This also implies that the extreme CAPES06 or EHI03 used may be not broadly applicable to identify SLS environments over the Southeast, as they in general capture high-CAPE environments while miss high-shear low-CAPE cases (Figure \ref{fig_era5_jointpdf}), and thus the composite synoptic patterns for R6 may be less representative of the SLS events in the Southeast. Sufficient low-level moisture supply is clearly essential for significant SLS environments, as all composite analyses reveal robust moisture transport at low levels. Deeper composite analysis using other composite methods such as empirical orthogonal functions \cite[EOFs;][]{schaefer1984}, rotated principal component analysis \cite[RPCA;][]{jones2004, Mercer_etal_2012}, or self-organizing maps \cite[SOMs;][]{sheridan2011} would be a valuable path for future work.

\section{Conclusions}\label{sec:conclusions}

This work provides a comprehensive climatological analysis and evaluation of SLS environments, and the associated synoptic-scale features that frequently generate them, in the ERA5 reanalysis data and CAM6 climate model simulation. Unlike reanalysis datasets, climate models, including CAM6, do not reproduce the observed daily weather, but they both are able to capture statistical states (e.g., the mean or extremes) for a  climate. Thus, we analyzed the overall climatology, as well as seasonal and diurnal cycles, of SLS environments and the occurrence frequency of the synoptic-scale features in the ERA5 reanalysis and a CAM6 AMIP-style simulation for years of 1980--2014 over North America. Here, SLS environments are measured by extreme values (defined as the 99th percentile) of two environmental proxies for SLS favorability, CAPES06 (the product of CAPE and S06) and EHI03 (proportional to the product of CAPE and SRH03), and their constituent parameters (CAPE, S06, and SRH03). Key synoptic-scale features commonly associated with the generation of SLS environments analyzed in this work include southerly Great Plains low-level jets, drylines, elevated mixed layers, and extratropical cyclone activity. Biases in these SLS environments and synoptic-scale features from CAM6 simulation were attributed to biases in the mean-state. Finally, composite analysis was conducted for six sub-regions over the eastern half of the United States to characterize the common synoptic patterns associated with significant SLS environments and assess their CAM6 model representation as compared to ERA5 reanalysis. Primary results are summarized as follows:

\begin{enumerate}
\item ERA5 reanalysis reasonably reproduces the observed (radiosonde) spatiotemporal distribution of SLS environments, with relatively low biases and strong correlations particularly over the Great Plains. Kinematic parameters (S06 and SRH03) are in general better estimated by ERA5 reanalysis than thermodynamic parameters (CAPE), especially for stations in the Mountain west and along the east coast where terrain effects are likely large.

\item Climatological patterns of extreme SLS environments over North America are reasonably well-captured by the ERA5 reanalysis and CAM6 simulation. Both ERA5 and CAM6 representations of extreme CAPES06 and EHI03 indicate qualitatively similar annual, seasonal, and diurnal climatologies of extreme SLS environments. Local maxima are found over southern Texas in spring and shift to the central United States in summer. The diurnal cycle peaks during the late afternoon and early evening with a minimum in the early morning, with larger amplitude over the continental interior and smaller amplitude in coastal regions. Extreme values of CAPES06 or EHI03 typically consist of very high CAPE and moderate-to-small S06 or SRH03, and thus the climatological behavior of these proxies are dominated by the behavior of CAPE extremes, not S06 or SRH03 extremes. This implies that extreme CAPES06 and EHI03 is less representative of high-shear, low-CAPE environments which contribute to a considerable portion of SLS environments in the Southeast.

\item Climatologies of key synoptic-scale features over North America are reasonably captured by the ERA5 reanalysis and CAM6 simulation as well. Southerly Great Plains low-level jets, drylines, elevated mixed layers, and extratropical cyclone activity in both ERA5 and CAM6 are most frequent east of the Rocky Mountains in warm seasons. Both ERA5 and CAM6 capture the strong linkage between SLS environments and these synoptic-scale features, as the spatial pattern and seasonal variation of the occurrence frequency of these synoptic-scale features are highly consistent with that of SLS environments. 

\item Biases between the CAM6 simulation and ERA5 reanalysis over the eastern United States are largest during summer: (1) CAPE extremes are biased high in CAM6 over much of the eastern half of the United States, which is primarily attributed to the enhanced surface specific humidity in CAM6; (2) SRH03 extremes are biased slightly high, which is primarily attributed to the stronger mean-state low-level winds in CAM6 than ERA5 and more frequent southerly Great Plains low-level jets; (3) elevated mixed layer frequency is biased high, which is primarily attributed to a steeper mean-state mid-level lapse rate and further enhances CAPE; and (4) taken together, the combined proxies CAPES06 and EHI03 are each biased high.

\item The composite synoptic patterns favorable for extreme SLS environments within six sub-regions across the eastern United States in the ERA5 reanalysis indicate an intensified upper-level jet streak in the vicinity of the region, sufficient warm and dry air advection from the elevated terrains eastward toward the Great Plains, and robust low-level moisture transport from the Gulf of Mexico due to the enhanced prevailing southerly or southwesterly winds east of a surface trough or cyclonic anomaly. CAM6 successfully reproduces these structures found in ERA5 for most regions. The principal exception is over the southeastern United States in CAM6 where the influence of the North Atlantic Subtropical High is also important, highlighting differences in climatological forcing that may translate to different behavior and predictability of severe weather itself, as has been identified in past work. 
\end{enumerate}

These results suggest that the ERA5 reanalysis data reasonably reproduce SLS environments and synoptic-scale features, and climate models such as CAM6 can be useful tools to investigate climate controls on the generation of SLS environments over North America. Meanwhile, it is necessary to be aware of the biases in climate models (e.g., the systematic biases in surface moisture and temperature over the central and eastern United States), which may affect the interpretation of their projections. To further understand the formation of SLS environments within climate system, future work can use idealized climate modeling experiments to quantitatively test both the detailed linkages between key synoptic-scale features and SLS environments, as well as how climate-scale boundary forcing fundamentally controls the spatiotemporal distribution of SLS environments on Earth.

\acknowledgments{ 
The authors thank Larry Oolman (University of Wyoming) for providing radiosonde data. The authors thank Editor Dr. Xin-Zhong Liang and three anonymous reviewers for their feedback in improving this manuscript. The authors would like to acknowledge high-performance computing support from Cheyenne (doi:10.5065/D6RX99HX) provided by NCAR's Computational and Information Systems Laboratory, sponsored by the National Science Foundation, for the simulation and data analysis performed for this work. This research was also supported in part through computational resources provided by Information Technology at Purdue, West Lafayette, Indiana. Li and Chavas were supported by NSF grant 1648681. Reed was supported by the NSF grant AGS1648629, as well as the U.S. Department of Energy Office of Science (DE-SC0019459). 
}

%






%
%
%
\bibliographystyle{ametsoc2014}
\newcommand{\noopsort}[1]{} \newcommand{\printfirst}[2]{#1}
  \newcommand{\singleletter}[1]{#1} \newcommand{\switchargs}[2]{#2#1}

%

%

\end{document}